\documentstyle[preprint,prd,aps,eqsecnum]{revtex}

\begin{document}

\preprint{{\vbox{\hbox {Jan. 1998} \hbox{IFP-741-UNC} \hbox{UR-1507}}}}

%\twocolumn[\hsize\textwidth\columnwidth\hsize\csname@twocolumnfalse\endcsname
%\draft

\title{A Solution to the Strong CP Problem with Gauge-Mediated Supersymmetry
Breaking}
\author{\bf Otto C. W. Kong$^{a,b}$
and Brian D. Wright$^a$
\footnote{Current E-mail address: bdwright@phy.ucsf.edu}}
\address{$^a$
Institute of Field Physics, Department of Physics and Astronomy,\\
University of North Carolina, Chapel Hill, NC  27599-3255\\
$^b$  Department of Physics and Astronomy,\\
University of Rochester, Rochester NY 14627-0171
\footnote{Present address; E-mail: kong@pas.rochester.edu} }
%\date{\today}
\maketitle

\begin{abstract}
We demonstrate that a certain class of low scale supersymmetric ``Nelson-Barr''
type models can solve the strong and supersymmetric CP problems while
at the same time generating sufficient weak CP violation in the
$K^{0}-\bar{K}^{0}$ system.
In order to prevent one-loop corrections to $\bar{\theta}$ which violate
bounds coming from the neutron electric dipole moment (EDM), one needs a scheme
for the soft supersymmetry breaking parameters which can naturally
give sufficient squark degeneracies and proportionality of
trilinear soft supersymmetry-breaking parameters to Yukawa couplings.
We show that a gauge-mediated supersymmetry breaking sector can provide
the needed degeneracy and proportionality, though that proves to be a problem
for  generic Nelson-Barr models. The workable model we consider here
has the Nelson-Barr mass texture enforced by a gauge symmetry; one also expects 
a new U(1) gauge superfield with mass in the TeV range. The resulting model is 
predictive. We predict a measureable neutron EDM and
the existence of extra vector-like quark superfields which can be discovered
at the CERN Large Hadron Collider.
Because the $3\times 3$ Cabbibo-Kobayashi-Maskawa matrix is approximately real,
the model also predicts a flat unitarity triangle and the absence of
substantial CP violation in the $B$ system at future $B$ factories.
We discuss the general issues pertaining to the construction of such a
workable model and how they lead to the successful strategy. A detailed
renormalization group study is then used to establish the feasibility
of the model considered.
\end{abstract}

%\vskip0pc]
%\vskip2pc]

\newpage

\section{Introduction}

The strong CP problem is without question one of the most
important problems faced by the Standard Model (SM). Its origin
lies in the necessity of adding the so-called $\theta$ term to the
effective QCD Lagrangian due to the contribution of instantons
present in the topologically nontrivial QCD vacuum\cite{instanton}:
\begin{equation}
{\mathcal L}_{{\mathrm eff}} = \frac{\theta\alpha_s}{8\pi}
F^A_{\mu\nu}\tilde{F}^{A\mu\nu}~,\label{leffqcd}
\end{equation}
where the dual field strength is given by $\tilde{F}_{\mu\nu} = \frac{1}{2}
\epsilon_{\mu\nu\alpha\beta}F^{\alpha\beta}$.
Through the anomaly in the axial ${\mathrm U}(1)$ current of QCD,
chiral U(1) transformations lead to shifts in $\theta$, leaving
the physical combination $\bar{\theta} = \theta - {\mathrm arg\, det} M_q$, where
$M_q$ is the quark mass matrix. Since ${\mathcal L}_{\mathrm eff}$ clearly 
violates CP, it gives a strong interaction contribution to the neutron 
electric dipole moment\cite{nedm} and leads to the experimental contraint
\begin{equation}
\bar{\theta} <   10^{-9}~.
\end{equation}
The real problem therefore is one of naturalness or fine-tuning:
Why is $\bar{\theta}$ so incredibly small?

There are currently three notable classes of possible solutions to this
problem: (1) vanishing up quark mass, (2) the axion\cite{visax,invisax}
or (3) CP conservation and subsequent spontaneous breaking.
The first and simplest possibility appears to be disfavored by
current algebra relations between pseudoscalar meson masses\cite{Weinberg},
but is still controversial (see {\it e.g.} Ref.~\cite{KapMan}).
Of these, the most popular is the invisible axion alternative\cite{invisax}.
Here one introduces a global chiral ${\mathrm U}(1)_{PQ}$
(Peccei-Quinn\cite{PQ})
symmetry which is spontaneously broken at a high energy scale $f$ and explicitly
broken by instantons. The $\theta$ parameter is replaced by a dynamical
field --- a pseudo Goldstone boson of the ${\mathrm U}(1)_{PQ}$ ---
whose potential dynamically relaxes $\bar{\theta}$ to zero.
The advantage of this scheme is that it is simple, generic and has
observable consequences both in terrestrial experiments and in
astrophysics and cosmology. Astrophysical constraints from axion-induced
cooling during stellar evolution\cite{Raffelt} and
effects on the neutrino signal from supernova 1987A\cite{Turner}
give a lower bound, $f \agt 10^{10}$ GeV, while a cosmological upper bound of
$\sim 10^{12}$ GeV is given by the contribution to the universal
energy density of the vacuum energy associated with
${\mathrm U}(1)_{PQ}$ breaking as the axion vacuum expectation value
relaxes to zero\cite{axcosmo}.
On the aesthetic side, one may complain that
we are merely replacing the $\bar{\theta}$ fine-tuning problem with another:
the smallness of the ratio of the weak scale to the ${\mathrm U}(1)_{PQ}$
breaking scale $\sim 10^{-(8{\mathrm -}10)}$.
Another possible problem is the dependence of the solution on a
global symmetry, generally not preserved by
gravity, so not likely to appear from a more fundamental theory.
This appears to be a significant problem\cite{aPl},
at least in Einstein gravity. However, it has been argued that
gravitational violations of global symmetries may be suppressed in
certain extensions, including string theories\cite{KLLS}, where at least the
universal dilaton-axion is always present.
The axion alternative also does not provide an explanation of
weak CP violation. Here one must assume the Kobayashi-Maskawa (KM) origin
of CP violating phases. Of course the ultimate test is to detect actually
an axion\cite{axexpts}.

In this paper we will focus on the third alternative. That is,
we will assume that the fundamental theory of nature preserves CP
and that at sub-Planck energies it is spontaneously broken. Indeed
there is evidence that CP has its origin as a gauge symmetry remnant
of superstring theories\cite{DLM}.
In this manner the smallness of $\bar{\theta}$ reflects the existence of an
underlying symmetry. Such models were first constructed in the context
of Grand Unified theories (GUTs) by
Nelson and refined by Barr\cite{NB} and incorporate extra heavy
quarks which mix with the observed quarks.
Then, relying on specific symmetries,
one can obtain a texture of the full quark mass matrices which guarantee
the tree-level vanishing of $\bar{\theta}$ after the spontaneous
CP violation (SCPV). After integrating out the heavy fields, the low
energy quark mass matrices contain the usual KM phase.
The generic difficulty with these models comes from the need to ensure that
large contributions to $\bar{\theta}$ do not arise at higher loops,
while at the same time having sufficient weak CP violation from the KM
phase. Thus while the SCPV approach is conceptually
rather simple, it is not so generic and requires careful model building.

Given some of the tantalizing hints of low energy supersymmetry
and the plausible gauge origin of CP symmetry, it is worthwhile to
attempt to construct  a supersymmetric (SUSY) model
with a Nelson-Barr type mechanism for solving the strong CP problem.
The SCPV feature  then also  resolves the so-called
SUSY phases problem. The latter problem was
originally described in the context of a minimal supergravity
origin of the soft SUSY breaking terms\cite{PW}, and is usually worse
in a general SUSY breaking scenario. There  are two extra
phases in the universal soft
mass parameters, beyond $\delta_{KM}$ and $\bar{\theta}$, which give
CP violating effects in the low energy effective theory.
These can be written as effective phases in the coefficients $A$ and
$B$ of the trilinear and bilinear soft SUSY breaking scalar terms,
respectively, given by
\begin{equation}
\phi_A = {\mathrm arg}(A M_{1/2}^\ast)\qquad
\phi_B = {\mathrm arg}(B M_{1/2}^\ast)~,\label{susyphases}
\end{equation}
where $M_{1/2}$ is the universal gaugino mass. The problem is
that from 1-loop diagrams involving squarks, these phases must be
fine-tuned to order $10^{-2}$ -- $10^{-3}$ to satisfy the limit on the
neutron electric dipole moment unless all the superpartners are
``heavy'', $\sim 1$ TeV. With CP  spontaneously broken in a sector
independent of SUSY breaking, these phases would be naturally zero
at first order.

Attempts to realize the Nelson-Barr mechanism in SUSY
models\cite{BMS} have, however, run up against a formidable
difficulty: There generically exist potentially
large 1-loop contributions to $\bar{\theta}$ in these models\cite{DKL}.
The dangerous diagrams are shown in Fig.1, where now in the
supersymmetric case $\bar{\theta}$ also gets contributions from the
argument of the gluino mass (Fig.1b):
\begin{equation}
\bar{\theta} = \theta - {\mathrm arg\, det} M_q - 3\, {\mathrm arg} M_g~.
\label{thetabarsusy}
\end{equation}
As discussed at length in Ref.~\cite{DKL}, one requires an
exceptionally high degree
of proportionality of the soft SUSY breaking trilinear scalar couplings
to their associated Yukawa couplings as well as
degeneracy among the soft squark mass terms for each charge and color
sector, if these contributions are to be sufficiently suppressed.
This is equivalent to the statement that when the quark and squark mass
matrices are diagonalized by the same set of unitary matrices, no
phase can appear in the diagrams of Fig.1. The degree
of proportionality and degeneracy required among the
soft SUSY breaking parameters is very difficult to maintain
due to the effects of renormalization.

So, is SCPV doomed to be disfavored as a solution to the strong CP problem
in supersymmetric models?
We will argue that models with a specifically modified Nelson-Barr
mechanism together with  the recently popular gauge-mediated
SUSY breaking (GMSB) scenario\cite{GMSB,MMM,fx,Bor,rev}
can overcome the difficulty. The  GMSB scenario ensures that
the soft masses at the intrinsic SUSY breaking scale $M_{mess}$ are
proportional and degenerate, while renormalization effects that violate
these conditions are reduced by having to run soft masses
and couplings from the messenger scale $M_{mess} \simeq 10 - 100$ TeV
instead of the reduced Planck mass $\simeq 2\times 10^{18}$ GeV.
In the minimal version of such GMSB models\cite{MMM,Bor}, the $A$- and $B$-terms 
are zero at $M_{mess}$. This tends to give additional
suppression of the dangerous contributions to $\bar{\theta}$.
However, large third generation Yukawa couplings can still lead to
significant violations of
proportionality and degeneracy and a detailed numerical analysis of
the situation is necessary to determine if the supersymmetric
Nelson-Barr type models are viable solutions to the strong CP problem.
We will answer this in the affirmative with what to our knowledge is the
only full renormalization group (RG) analysis of such models in the literature.

The models to be discussed here have CP spontaneously broken at low energies 
(of order a TeV), and a source of weak CP violation distinct from that in the 
Standard Model, namely the exchange of a new U(1) gauge boson, the symmetry
of which enforces the Nelson-Barr texture. The KM phase is very small.
This makes it possible to account
for the smallness of $\bar{\theta}$ without making weak CP violation
inconsistent with observations. In the SUSY-GUT Nelson-Barr models,
even with gauge-mediated SUSY breaking, one predicts a too
large $\bar{\theta}$ if the experimental requirement that $\delta_{KM}
\sim {\mathcal O}(1)$ is imposed. Moreover, one has a richer and quite distinct 
phenomenology. The extra quarks and other fields needed to construct the 
Nelson-Barr texture and break CP will be within the reach of future accelerators.
A non-supersymmetric version of the type of model is the aspon
model\cite{aspon,FN,asponB}. The situation with SUSY incorporated is first 
discussed in Ref.\cite{FK} where its advantage over the generic SUSY
Nelson-Barr type models is highlighted.
Another possible advantage for such low scale models would arise if it were
somehow possible to imbed the sector responsible for the Nelson-Barr
texture in the SUSY breaking and messenger sectors.
An immediate disadvantage of this approach is that by breaking CP
at low energies, one introduces a serious domain wall problem\cite{domwall}.
However, this can be solved via a period of inflation just above
the weak scale. Indeed, in the SUSY context, some authors
have argued that this type of inflation can be natural\cite{weakinfl}
and desirable for other reasons ({\it e.g.} as a solution to the cosmological
moduli problem).

This article is organized as follows: In Section 2,  the considerations leading to 
viable models for solving the strong and supersymmetric CP problems are discussed.
We also note some intriguing alternatives worthy of further consideration. We give 
in Section 3 a detailed summary of the dangerous 1-loop contributions to 
$\bar{\theta}$ and their dependence on proportionality and squark-degeneracy 
violating mass insertions. The question of the detailed
structure of the spontaneous CP breaking part of the superpotential
is taken up in Section 4. We emphasize its importance in determining certain 
dangerous contributions to $\bar{\theta}$ and present a minimal example we use
in further analysis. The renormalization group analysis used to estimate the 
$\bar{\theta}$ contributions is described in Section 5 and the
full numerical results presented. Some remarks on related questions
of interest are presented in the conclusion.

\section{Model-Building Considerations}

We assume  full CP symmetry
in the visible sector, including the soft SUSY breaking terms,
down to energies where spontaneous CP breaking occurs.
From Eq.(\ref{thetabarsusy}) we see that there should be no
tree-level phases in the quark mass determinant, nor in the SUSY
breaking gluino mass. The latter holds by assumption and the former
is obtained through a Nelson-Barr texture. To obtain the texture,
we introduce an extra heavy
right-handed down quark superfield $\bar{D}$ together with its mirror
$D$ coupling to the ordinary down quarks via the superpotential\footnote{
Other type of phenomenological features from an extra vector-like quark
singlet have been studied by various authors in a different context. See
for example Ref.\cite{singlet}, and references therein.}
\begin{equation}
W_d = Y_d^{ij}Q_j\bar{d}_i H_d + \mu_{{\scriptscriptstyle D}} D\bar{D} +
\gamma^{ia}D\chi_a \bar{d}_i~,\label{Wd}
\end{equation}
where the VEVs of the scalar components of $\chi_a$ contain a relative
phase, thus breaking CP. The details of the superpotential accomplishing
this are postponed to Section 4, and here it is sufficient to note that
at least two $\chi$ fields are necessary, since one phase can always be
absorbed by a field redefinition of the extra quarks. After CP and
${\rm SU}(2)_W\times {\rm U}(1)_Y$ breaking, we have the down sector fermion
mass matrix:
\begin{equation}
m_q = \left( \begin{array}{cc}
m_d & x  \mu_{{\scriptscriptstyle D}} \mbox{\boldmath $a$} \\
0 & \mu_{{\scriptscriptstyle D}}
\end{array}\right)~,\label{mf}
\end{equation}
where $m_d$ is the usual $3\times 3$ down sector mass matrix
and {\boldmath $a$} is a complex $3$-vector with components
$a^i = \frac{1}{x \mu_{{\scriptscriptstyle D}}}
\gamma^{ia}\langle\chi_a\rangle$, and the real parameter $x$  is defined 
such that {\boldmath $a$}  is normalized to 1, {\it i.e.}
{\boldmath $a^{\dag} a$}$=1$. The magnitude of mixing between the ordinary
quarks and the extra singlet is characterized by $x$. Clearly
the determinant of $m_q$ is real
and at energies below $\mu_{{\scriptscriptstyle D}}$ the low energy
effective theory has a KM phase of at most order $x$. Without some
additional source for weak CP violation, this must be ${\mathcal O}(1)$
and as we shall see this in turn makes the suppression of 1-loop
contributions to $\bar{\theta}$ problematic.\footnote{This is exactly the
reason why the small $x$ scenario is discarded in the analysis
of Ref.\cite{DKL}.} For this reason, we shall
assume that CP is broken at relatively low scales
with a nonstandard mechanism for weak CP violation.

The specific form of the mass matrix in Eq.(\ref{mf}) can be enforced
by a variety of symmetries, though by the non-renormalization theorems,
the Nelson-Barr texture is not upset by renormalization of terms in
the superpotential.  In the aspon scenario,  as discussed in
Ref.\cite{FN}, the $D$, $\bar{D}$ and $\chi_a$ can be given charges 
$1,-1$ and $1$, respectively, under a new gauged U(1) symmetry.  The major 
source of weak CP violation, in the $K-\bar{K}$ system for instance, then
comes from exchange of the new U(1) gauge boson (aspon)
which becomes massive at the scale where CP is spontaneously
broken. This places  an upper bound on the mass scale of CP
breaking of ${\mathcal O}({\rm TeV})$. More important in our SUSY
version, this allows the parameter $x$ to be small, {\it e.g.}
$x^2\sim 10^{-5}$, which contributes significantly in suppression 
$\bar{\theta}$ from loop corrections. Note that we will need at least some 
extra mirror partners for $\chi_a$ superfields to cancel the gauge 
anomaly introduced by their fermionic components.

If the Nelson-Barr texture is obtained from a discrete/global symmetry,
one must rely on superbox diagrams involving gluino and chargino exchange 
to generate $\epsilon_K$; a scenario recently re-analyzed in the context of 
the minimal supersymmetric Standard Model (MSSM)\cite{epK}.
However, in present setting, this proves very difficult to do.
This is unfortunate since it is easier to construct unifiable models
in this latter case.\footnote{To construct a unifiable model in the discrete
symmetry case, we need to add a pair of heavy lepton doublets
coupling in a superpotential analogous to Eq.(\ref{Wd}). The
same thing can be done to make the aspon model GUT-compatible,
but unifying the extra U(1) with the other gauge interactions
is not  possible.}

When SUSY is broken we generate nonzero gaugino masses and soft scalar bilinear 
and trilinear couplings,\footnote{Note that the trilinear couplings,
$h_d$ and $h^{ia}_{\gamma}$,
are also commonly written as  products of the $A$-parameters and
the corresponding Yukawa couplings, {\it e.g.} elements of the
$h_d$ matrix correspond to $A_d^{ij}Y_d^{ij}$ (no sum).} including
the following terms relevant for the down-sector analysis [a complete
description is given in Eqs.(\ref{mssmsoft},\ref{extrasoft})]:
\begin{eqnarray}
V_{\text{soft}}^d & = & \hat{Q}^{\dagger}\tilde{m}_Q^2 \hat{Q}
+ \hat{\bar{d}} \tilde{m}_{\bar{d}}^2 \hat{\bar{d}}^{\dagger}
+ \hat{D}^{\dagger}\tilde{m}_D^2 \hat{D}
+ \hat{\bar{D}} \tilde{m}_{\bar{D}}^2 \hat{\bar{D}}^{\dagger}
+ \chi_a^{} \tilde{m}_{\chi ab}^2 \chi_b^{\dagger}
\nonumber \\
&&{} + \hat{\bar{d}} h_d \hat{Q} H_d
+ \hat{\bar{d}}_i h_{\gamma}^{ia} \hat{D} \chi_a
+ B_{{\scriptscriptstyle D}} \mu_{{\scriptscriptstyle D}}\hat{\bar{D}} \hat{D}
+ \text{h.c.} \label{vdsoft}
\end{eqnarray}
The general form of the down squark mass matrix can be written as,
\begin{equation}
{\mathcal M}^2_d = \left( \begin{array}{cc}
{\mathcal M}_{RR}^2 & {\mathcal M}_{RL}^2 \\
{\mathcal M}_{RL}^{2 \dagger} & {\mathcal M}_{LL}^2
\end{array} \right)~, \label{8sqm}
\end{equation}
where
\begin{eqnarray}
{\mathcal M}_{LL}^2 & = & \left( \begin{array}{cc}
\tilde{m}_d^2 + m_d^{\dag}m_d & x \mu_{{\scriptscriptstyle D}}
m_d^{\small \text{T}} \mbox{\boldmath $a$} \\
x \mu_{{\scriptscriptstyle D}} \mbox{\boldmath $a^\dagger$} m_d &
\mu_{{\scriptscriptstyle D}}^2 ( 1 + x^2) + \tilde{m}_D^2
\end{array}\right)~,\label{mll} \\
{\mathcal M}_{RR}^2 & = & \left( \begin{array}{cc}
\tilde{m}_{\bar{d}}^2 + m_d m_d^{\text{T}} +
x^2 \mu_{{\scriptscriptstyle D}}^2 \mbox{\boldmath $a a^\dagger$} &
x \mu_{{\scriptscriptstyle D}}^2 \mbox{\boldmath $a$} \\
x \mu_{{\scriptscriptstyle D}}^2 \mbox{\boldmath $a^\dagger$} &
\mu_{{\scriptscriptstyle D}}^2 + \tilde{m}_{\bar{D}}^2
\end{array}\right)~,\label{mrr} \\
{\mathcal M}_{RL}^2 & = & \left( \begin{array}{cc}
h_d v_d +  m_d \mu_{{\scriptscriptstyle H}} \tan\!\beta
 & M_5^2 \mbox{\boldmath $b$} \\
0 & B_{{\scriptscriptstyle D}} \mu_{{\scriptscriptstyle D}}
\end{array}\right)~.\label{mrl}
\end{eqnarray}
In the expression for ${\mathcal M}_{RL}^2$ we have used
\begin{equation}
M_5^2 b^i = h_{\gamma}^{ia}\langle\chi_a\rangle - \gamma^{ia}
\langle F_{\chi_a}\rangle~,\label{mrloffdiag}
\end{equation}
where {\boldmath $b$} is normalized to 1 and $F_{\chi_a}$ is the $F$-term
for the $\chi_a$ field, which depends on the specific form of the
soft SUSY breaking mass terms related to the
spontaneous CP breaking part of the superpotential.
The form of these squark mass matrices will be critical in the calculation
of $\bar{\theta}$ in the next section. In particular, the
$\langle F_{\chi_a}\rangle$'s  bear complex phases
independent of those in the $\langle\chi_a\rangle$'s in a generic setting,
and hence constitute a major source of trouble.

We have implicitly assumed in the above that the sectors responsible for
the Nelson-Barr texture and CP breaking are disjoint from those involved
in the intrinsic breaking of supersymmetry. Since the successful example model
we focus on has gauge-mediated SUSY breaking at a relatively
low scale $M_{mess}$, it is {\it a priori} possible that all or part
of the extra field content and symmetries required for the Nelson-Barr
mechanism is contained in the SUSY breaking hidden and messenger sectors.
Although this is an intriguing possibility, we have not been able to
construct viable models of this type thus far. Actually, there appears to
be an intrinsic incompatibility between the role a field takes in
CP violation and the one it takes in SUSY breaking, as far as constraining
$\bar{\theta}$ is concerned. Recall that in GMSB, the scalar particles
get their soft SUSY breaking masses from the gauge interactions
they share with the messenger sector particles which see SUSY
breaking directly. The squark masses in each sector, for instance,
would then be degenerate, as the process is flavor blind.
The  extra singlet  $\bar{D}$ introduced here may easily upset the situation.
Naively, the best strategy is to make the GMSB  also blind to
the aspon U(1). We will see below that this happens to have
a even more important merit --- it guarantees the suppression of the
very dangerous $\left\langle F_{\chi_a}\right\rangle$'s.
In other words, hiding the CP-breaking sector from
SUSY breaking helps to suppress the SUSY loop contributions
to $\bar{\theta}$. This is the less ambitious strategy  we have taken
in the model analyzed in detail below.

\section{Calculation of 1-Loop $\bar{\theta}$ Constraints}

Here we review and extend the work in Refs.\cite{DKL,FK} to compute
the 1-loop contributions to $\bar{\theta}$ of Figs.1a and 1b, using
the mass-insertion approximation\cite{mia}. These results are generally
valid for any type of low energy SCPV model, with or without the
U(1)$_A$. The scale,  $\mu_{\scriptscriptstyle D}$, and hence the 
characteristic scale of the SCPV, here is chosen near or below 
$\tilde{m}_{sq}$, the average squark mass. This is more or less
dictated by the aspon scenario of weak CP\cite{FN}.
The $D$ and $\bar{D}$ superfields are handled
on the same footing as the other quark superfields. The analysis
is basically the same as that given in Ref.\cite{FK} except  here
we pay full attention to the explicit phase factors and family indices,
and also treat the $M_5^2$ term,
as given by Eq.(\ref{mrloffdiag}), in full detail. These turn out to be
very important in understanding how the scenario can provide a
feasible solution. In taking the large
$\mu_{\scriptscriptstyle D}$ limit, which corresponds to situation
discussed in Ref.\cite{DKL}, one has to be careful in handling
the loop momentum integrals properly. The latter are however not explicitly
given in this paper, though they are included in our numerical computations.

The 1-loop contribution is given by
\begin{eqnarray}
\delta\bar{\theta} &=& \text{Im Tr}\,
 m_{\scriptscriptstyle F}^{-1} \delta\! m_{\scriptscriptstyle F}
+ 3  M_g^{-1} \delta\!M_g \nonumber \\
&=& \frac{\alpha_s}{4\pi} \sum_{i,I} {\rm Im}[Z^{iI*}Z^{(i+4)I}]
{\mathcal M}^2_{d{\scriptscriptstyle I}}
\left( \frac{M_g}{m_{\scriptscriptstyle F\!i}}
\frac{8/3}{M_g^2-{\mathcal M}^2_{d{\scriptscriptstyle I}}}
\ln \frac{{\mathcal M}^2_{d{\scriptscriptstyle I}} }{M_g^2}
+ \frac{m_{\scriptscriptstyle F\!i}}{M_g}
\frac{3}{{\mathcal M}^2_{d{\scriptscriptstyle I}}
 - m_{\scriptscriptstyle F\!i}^2}
\ln \frac{{\mathcal M}^2_{d{\scriptscriptstyle I}} }{ m_{\scriptscriptstyle F\!i}^2}
\right) \; ,
\end{eqnarray}
where $m_{\scriptscriptstyle F\!i}$ runs over the four eigenvalues of
the quark mass matrix [cf. Eq.(\ref{mf})] and
${\mathcal M}^2_{d{\scriptscriptstyle I}}$
over the eight eigenvalues for the squarks [cf. Eq.(\ref{8sqm})],
all in the down-sector; $Z^{IJ}$ is the unitary rotation that diagonalizes
the squark mass matrix in the quark mass eigenstate basis. This full
formula, while it can be used in the numerical calculations once all the
quantities involved are known, hides its physics content behind
the $Z$-matrix elements. In the limit of exact degeneracy and
proportionality, the latter is just the identity matrix and $\bar{\theta}$
is zero. Otherwise, the mass-insertion approximation, as discussed
below, is more illustrative.

We first assume an approximate degeneracy and that the diagonal
blocks in ${\mathcal M}_d^2$ dominate over the off-diagonal
block ${\mathcal M}^2_{RL}$ and write
\begin{eqnarray}
\tilde{m}^2_{\bar{d}} =   \bar{m}^2_{\bar{d}} \times 1\!\!\!1
+ \delta\tilde{m}^2_{\bar{d}} \; ,
\quad \tilde{m}^2_{\bar{\scriptstyle D}} =   \bar{m}^2_{\bar{d}}
+ \delta\tilde{m}^2_{\bar{\scriptstyle D}} \; , \nonumber \\
\tilde{m}^2_{d} =   \bar{m}^2_{d} \times 1\!\!\!1
+ \delta\tilde{m}^2_{d} \; ,
\quad \tilde{m}^2_{{\scriptstyle D}} =   \bar{m}^2_{d}
+ \delta\tilde{m}^2_{{\scriptstyle D}} \; .
\end{eqnarray}
The squarks are then treated as scalars of masses
$\bar{m}^2_{\bar{d}}$ and $\bar{m}^2_{d}$ with the
$\delta\tilde{m}^2_{..}$ and ${\mathcal M}^2_{RL}$ treated
as admissible mass-insertions is the loop-diagrams Figs.1a and 1b.
Explicit forms of the matrices needed to diagonalize $m_{\scriptstyle F}$
are useful. Expressions up to order $x^2$ are available in the
literature\cite{x2}. To parametrize the effect of proportionality violation
among the three families, we write
\begin{equation}
h_d = \bar{A}_d Y_d + \delta\!A_d \; .
\end{equation}
The situation for the related parameter in the $d$-$D$ mixings is more complicated.
Recall that $M_5^2 b^i = h_{\gamma}^{ia}\langle\chi_a\rangle - \gamma^{ia}
\langle F_{\chi_a}\rangle$ [Eq.(\ref{mrloffdiag})].
It has been emphasized in Ref.\cite{FK} that  the $F$-terms
being small is paramount to the success of any model of the Nelson-Barr type.
These terms are dangerous because in general one
has no reason to expect these $F$-terms to obey even an approximate
proportionality (to the $x \mu_{\scriptscriptstyle D} a^i$ terms).
On the contrary, contributions of the other part to $\bar{\theta}$ can be 
interpreted as a proportionality violation among the $\gamma^{ia}$'s  by writing
\begin{equation}
h^{ia}_{\gamma}  = \bar{A}_{\gamma} \gamma^{ia}
 + \delta\!A_{\gamma}^{ia}\; ;
\end{equation}
the term proportional to $\bar{A}_{\gamma}$ does not contribute.
We further introduce the simplified notation:
\begin{equation}
\delta\!A_{\gamma} c^i = \frac{1}{x \mu_{\scriptscriptstyle D}} h^{ia}_{\gamma} 
\left\langle \chi_a \right\rangle - \bar{A}_{\gamma}  a^i~,  	\label{dagam}
\end{equation}
where complex vector {\boldmath $c$} is normalized to 1. Hence,
we have
\begin{equation}
M_5^2 b^i =  \bar{A}_{\gamma}( x  \mu_{\scriptscriptstyle D}  a^i )
+ \delta\!A_{\gamma} ( x  \mu_{\scriptscriptstyle D} c^i ) - \gamma^{ia}
\left\langle F_{\chi_a}\right\rangle~. \label{mfiveb}
\end{equation}

In terms of the above notation, the list of major contributions to
$\bar{\theta}$ is given in Tables 1a and 1b. The $\bar{\theta}$ contributions
involving  $M_5^2$ are complicated. To make it easier
to  see the effects of the different parts, we list some of those
terms in tables before and after the above mentioned splitting. For example,
entry 1 in the Table~1a is split into two parts: the first part is
a proportionality violation effect involving $\delta\!A_{\gamma}$ and
Im($a^*_ic^i$) (both are suppressed in our model), the second is the
$F$-term contribution ($\gamma^{ia} \left\langle F_{\chi_a}\right\rangle$),
where the relevant complex phase is taken to be ${\cal O}(1)$. One other notable 
feature among the $\bar{\theta}$ contributions is the combination
$M_5^2 b^i - x  \mu_{\scriptscriptstyle D} B_{\scriptscriptstyle D} a^i$,
as shown in entry 9 of Table~1b. When the $M_5^2 b^i$ term is split as above, 
the second term actually can be combined with the first term in Eq.(\ref{mfiveb}) 
to give $x  \mu_{\scriptscriptstyle D} (\bar{A}_{\gamma} -
 B_{\scriptscriptstyle D})  a^i$, which can be interpreted as a proportionality
violation  among the corresponding trilinear and bilinear
terms. The other parts involve  $\delta\!A_{\gamma}$ and
$\gamma^{ia} \left\langle F_{\chi_a}\right\rangle$, as explicitly
shown in the table. All other entries with a $M_5^2$ can be split
and interpreted in the same way. We will see in the final result that
the $F$-term contribution {\it is the most dangerous}.

\section{The Spontaneous CP Violation Sector}

Spontaneous breaking of the U(1)$_A$ symmetry is the only source
of CP violation in our model. This CP violation effect feeds
directly into the $x \mu_{\scriptscriptstyle D}$ and $M_5^2$ terms in
the quark and squark mass matrices, with complex phase vectors
{\boldmath $a$} and {\boldmath $b$}, respectively. To implement the
mechanism, we need a sector of U(1)$_A$-charged SM singlet
superfield with a superpotential that not only gives rise to the complex
$\left\langle \chi_a \right\rangle$'s, but also gives us a good control on the
dangerous $\left\langle F_{\chi_a} \right\rangle$'s. Soft SUSY breaking terms
should also be taken into consideration, when determining the true
scalar potential. The $F$-terms, of course, characterize
SUSY breaking. We consider the scenario in which the
messengers communicating SUSY breaking to the visible sector are
U(1)$_A$-blind, {\it i.e.} they do not carry any U(1)$_A$ charges;
furthermore, they are not directly coupled to the CP-breaking sector.
The superfields of the latter are then hidden from SUSY breaking.

We have to consider at least five superfields, two $\bar{\chi}$'s
of conjugate U(1)$_A$ charges to the $\chi_a$'s and a singlet
$\aleph$, in order to have both gauge anomaly cancellation and a possible
CP violating vacuum solution\cite{FK,4h}. We consider the
superpotential\footnote{In Ref.\cite{FK}, a $W_{\chi}$ without
the $\mu_{\chi}$-terms is suggested.  While that could have
a CP violating vacuum with SUSY preserved, the situation is
not as general and natural as the one considered here, and would
have to rely on a linear $\aleph$ term to fix the symmetry breaking scale.}
\begin{equation}
W_{\chi} =  \bar{\chi}_a \mu_{\chi}^{ab} \chi_{b}
+ \aleph \bar{\chi}_a \lambda^{ab} \chi_b
+ \lambda_{{\scriptscriptstyle \aleph}} \aleph^3
+ \mu_{{\scriptscriptstyle \aleph}} \aleph^2 \label{wchi} \; .
\end{equation}
The five $F$-flat conditions yield four independent equations,
which, together with the $D$-flat condition, give a unique vacuum solution.
The solution is CP violating for most  of the parameter space.
Hence, neglecting the soft SUSY breaking terms, we have a
SUSY preserving vacuum that breaks CP.

The GMSB scenario we considered allows the unwanted
soft SUSY breaking terms of the sector to be zero at $M_{mess}$.
They are, however, generated through RG evolution, as discussed
in the next section. With their nonvanishing values taken into
consideration, the scalar potential is then given by
\begin{equation}
V_{\chi} = D_{\chi}^2 + F^{}_{\chi_a}  F_{\chi_a}^*
+  F^{}_{\bar{\chi}_a} F_{\bar{\chi}_a}^* + F^{}_{\scriptscriptstyle \aleph}
F_{\scriptscriptstyle \aleph}^* + V_{s\chi}
\end{equation}
where
\begin{eqnarray}
V_{s\chi} &=&
 \bar{\chi}_a B_{\chi}^{ab} \chi_b
+ \bar{\chi}_a h_{\lambda}^{ab} \chi_b \aleph
 + h_{\scriptscriptstyle \aleph} \aleph^3
+ B_{{\scriptscriptstyle \aleph}}\aleph^2 \nonumber \\
& &
+ \bar{\chi}^{\dagger}_a \tilde{m}_{\bar{\chi} ab}^2
\bar{\chi}_b^{}
+ \chi_a^{} \tilde{m}_{\chi ab}^2 \chi_b^{\dagger}
+ \aleph^{\dagger} \tilde{m}_{\scriptscriptstyle \aleph}^2 \aleph \; .
\end{eqnarray}
Solving for the potential minimum to determine
the $\left\langle F_{\chi_a}\right\rangle$ values is not tractable, as
$W_{\chi}$ and  $V_{s\chi}$ involves a large number of parameters which are not
otherwise constrained, apart from yielding a CP violating solution.
However, one can easily obtain a reasonable order of magnitude estimate of the
shifts in  the $\left\langle F_{\chi_a}\right\rangle$'s as a result of 
including the small $V_{s\chi}$ terms. For example, the equation
\begin{equation}
\frac{\partial V_{\chi}}{\partial \bar{\chi}_a}
= 2D_{\chi}\frac{\partial D_{\chi}}{\partial \bar{\chi}_a}
- ( \mu_{\chi}^{ab}  + \aleph  \lambda^{ab} ) F_{\chi_b}
+ F_{\scriptscriptstyle \aleph} \frac{\partial
F_{\scriptscriptstyle \aleph}^*}{\partial \bar{\chi}_a}
+  B_{\chi}^{ab} \chi_b +  h_{\lambda}^{ab} \chi_b \aleph
+ \bar{\chi}^{\dagger}_b \tilde{m}_{\bar{\chi} ba}^2 =0
\end{equation}
suggests that $\left\langle F_{\chi}\right\rangle$ (here we drop all indices 
and phases) is given by the magnitude of
\begin{equation}
B_{\chi} \quad
\text{or} \quad \tilde{m}_{\bar{\chi}}^2  \quad\
\text{or} \quad   
 h_{\lambda} \left\langle \chi \right\rangle   / \lambda  \; .\label{Fest}
\end{equation}
An alternative way to estimate the $\left\langle F_{\chi_a}\right\rangle$'s
is given by
the SUSY breaking diagrams shown in Fig.2.
Here Figs.2a and 2b give exactly the same results as the first two
terms listed above. Figure~2c, however, gives the estimate
$\left\langle F_{\chi}\right\rangle \sim h_{\lambda}\left\langle \chi 
\right\rangle \lambda / 16\pi^2$. For perturbative
values of the $\lambda$ coupling, this is of course smaller
than the third estimate in Eq.(\ref{Fest}), hence we neglect it.
A similar diagram, Fig.2d, also suggests a contribution
$\sim h_{\gamma}\left\langle \chi \right\rangle \gamma / 16\pi^2$, though 
the $\gamma$ dependence of the $\left\langle F_{\chi} \right\rangle$'s is 
implicitly incorporated into the generation of the $V_{s\chi}$ terms
through RG running.  We will use all these in our numerical estimates to
determine whether the $F$-term is sufficiently small that its contributions
to $\bar{\theta}$, listed in Tables 1 and 2, are under control.
Finally, we emphasize again that the complex phases in
the $\left\langle F_{\chi_a}\right\rangle$'s are not related to those of the
$\left\langle \chi_a \right\rangle$'s directly.

\section{Renormalization Group Analysis}

As pointed out in the introduction, we need a full theory for the soft
SUSY breaking parameters to see if the $\bar{\theta}$ constraints
can be satisfied and the GMSB scenario may provide
the only viable possibility. In particular, we use here only the
minimal version of such a theory \cite{MMM}. This version has a few special
merits: it provides practically a one-parameter model of soft SUSY
breaking, radiative breaking of electroweak symmetry is naturally
implemented and, within the MSSM framework, it has been
studied with extensive renormalization
group analysis and shown to be compatible with all known experimental
constraints\cite{Bor}. From our perspective of solving the strong
CP problem by augmenting the  ${\mathrm{U(1)}}_A$ sector, it
actually represents a relatively demanding setting among GMSB
models,  where a large $\tan\!\beta$ allows all the Yukawa
couplings of the third family to have substantial effects on the
RG-runnings. A smaller  $\tan\!\beta$ in general would only make it easier
to satisfy the  $\bar{\theta}$ constraints.

We will refrain from elaborating extensively on the details of the GMSB model 
or the RG-analysis itself. For more specific details on finding the correct 
electroweak symmetry breaking vacuum and meeting other experimental
constraints in the minimal GMSB model, readers are referred
to Ref.\cite{Bor}. {\it Our interest here is
in adapting the machinery to our extended model at a level of
sufficient sophistication to calculate $\bar{\theta}$
to 1-loop and establish our solution to the strong CP problem.}

We use 1-loop renormalization group equations (RGE's) with naive
step thresholds between $M_Z = M_{SUSY}$ and $M_{mess}$.
The RG-improved tree level Higgs potential is considered in finding
the electroweak-symmetry breaking solution. The RGE's for the extra content of 
the model are also implemented at the 1-loop level; relevant formulae are
in Appendix A. The ${\mathrm{U(1)}}_A$,
and hence CP symmetry, breaking is imposed by hand. The idea is to study
the general situation independent of the details of the
SCPV sector, as the latter is to a certain extent more flexible and
less constrained.  It is important to note that the extra
superfield content in the model is partially decoupled from the MSSM
part, with the only direct coupling being gauge couplings of
$D$ and $\bar{D}$, and the small Yukawa couplings $\gamma^{ia}$.
The  computation concerning  the SCPV sector, as well as the RG analysis
can certainly be made more sophisticated, however, we consider
our treatment sufficient for our purpose.  In the sample analysis
for which numerical results are presented in this paper in detail
(Appendix~B  and the last column of Table~1), the values of the
various $\gamma^{ia}$ Yukawa couplings are generated randomly in
the range $0.005 - 0.01$. The latter is chosen to
target an $x$-value of around $0.01$. The value of
$\mu_{\scriptscriptstyle D}$ is fixed at $500$ GeV;
$M_{mess} \sim  \Lambda$ at $50$ TeV. For the soft SUSY breaking parameters
from GMSB, all $A$- and $B$- terms are taken to be zero at $M_{mess}\equiv X$.
The scalar soft masses from GMSB are given by
\begin{equation}
 {\tilde{m}^2(X)}  =    \frac{\Lambda^2}{8\pi}   \left\{
     C_3 \,\alpha_3^2(X) + C_2 \,\alpha_2^2(X)
           + \frac{3}{5} Y^2 \,\alpha_1^2(X)
                            \right\}\,f(y)    \; ,
\label{bcscal}
\end{equation}
where $C_3 = 4/3, 0$ for triplets and singlets of $SU(3)_C$,
$C_2 =3/4,0$ for doublets and singlets of $SU(2)_L$; $Y =Q-T_3$ is the
hypercharge. The function $f(y)$, derived in Ref.\cite{fx}, is simply set to 1.
Note that the above formula is independent of the  ${\mathrm{U(1)}}_A$
charge; a SM singlet with or without ${\mathrm{U(1)}}_A$ charge,
such as the $\chi_a$ and $\aleph$ scalars, has no initial soft mass.
Gaugino masses are  likewise given by the MSSM formula, omitted here.
The new  ${\mathrm{U(1)}}_A$ gaugino (aspino) has no tree-level SUSY breaking mass.
The gauge coupling $g_A$ is taken to be around $g_{em}$. The
SUSY-breaking aspino mass $M_A$ then remains vanishingly small even after
finite loop effects and RG evolution are taken into account.

After the electroweak symmetry breaking solution is obtained, various
$\bar{\theta}$ contributions are calculated through the
mass-insertion approximation to order $x^2$. To impose
the  ${\mathrm{U(1)}}_A$ symmetry breaking, we set, in the sample
analysis, $|\left\langle \chi_1 \right\rangle |^2 + |\left\langle \chi_2 
\right\rangle |^2 \simeq \mu_{\scriptscriptstyle D}^2$ and choose random values for
the VEVs and their complex phases within the constraint. Effects of higher
order in $x$  are  checked to be insignificant.

Values of parameters in $W_{\chi}$  are needed for the RG-runnings of 
particularly the SCPV sector soft SUSY breaking parameters, discussed in the
previous section. To simplify the situation, we input all these mass
parameters as $ \mu_{{\scriptscriptstyle D}}$ and all dimensionless
couplings as random numbers in the range $0.1-0.8$, for the sample
calculation. This oversimplification certainly
begs the question of consistency of the vacuum solution for this
sector, or the whole model. However, in the small $x$ domain
of interest, the influence of the extra ingredients on the values of
the other MSSM parameters is insignificant, as to be expected.
The only practical effect of those parameters is in the  RG evolution
of the related soft terms which we needed to estimate
the  $\left\langle F_{\chi} \right\rangle$'s. We have checked, for instance, that
the particular input values used in the sample run reported here
does lead to generic magnitudes of  the latter.

Appendix B contains
a collection of some of the numerical results, while those for the
$\bar{\theta}$ contributions, {\it without} the $\left\langle F_{\chi} 
\right\rangle$'s are listed in the last column of Table 1. Estimates of the
$\left\langle F_{\chi} \right\rangle$, following the discussion in the 
previous section, and their contribution to $\bar{\theta}$ are given in Table 2 
(second column). The  latter can be  easily checked using the 
$\left\langle F_{\chi} \right\rangle$ value and the listing of $M_5^2$-terms 
in Table 1. We also list in Table 2 results
from a number of different runs with different values of the $\gamma$'s
(reflected by the $x$-value obtained) and $\lambda$'s. The former, which
can have a significant effect on the various MSSM parameters, are restricted
by $x^2 \sim 10^{-3}$ -- $10^{-5}$ from the weak CP considerations. Our
results indicate that only a relatively large value of $x$ can upset
the strong CP solution, by first driving $\left\langle F_{\chi} \right\rangle$ 
too large (see column 4 of Table 2). One should be cautious in using this result
quantitatively, as our $\left\langle F_{\chi} \right\rangle$ estimates are meant 
to be conservative upper bounds. However, the result is certainly illustrative
of the importance of the $\left\langle F_{\chi}\right\rangle$ in estimating 
$\bar{\theta}$. With $x$ restricted to the workable range, the
basic features of the RGE solutions are quite stable. This is true even
with a relatively large variation of the $\lambda$'s, as illustrated
by column 5 and 6 of Table 2. Note that though the
$h_\lambda \left\langle \chi \right\rangle / \lambda$
term may have an explicit dependence on $\lambda$, its numerical value
does not have a large variation with $\lambda$ as
one might naively expect. This is, like the approximate proportionality
of a general $A$-term, a natural result of the RG equations.

All in all, the $F$-term contributions to $\bar{\theta}$ dominate,
and the overall $\bar{\theta}$ value is comfortably within the required
bound for the major region of the parameter space of our model under
consideration, hence solving the strong CP problem.

\section{Conclusions}

To recapitulate, we discussed a complete spontaneous CP violation
model with gauge-mediated supersymmetry breaking and why this type
of model is particularly favored over a generic supersymmetric
SCPV model in solving the strong CP problem.
Results from  numerical RGE studies are used to explicitly establish
the feasibility of the approach. The treatment of parameters in the SCPV
sector is admittedly oversimplified. The correlations between $x$ and
$\mu_{\scriptscriptstyle D}$, and between the various mass parameters
at the  $\mu_{\scriptscriptstyle D}$ scale and the values of the
various $\lambda$ coupling, for instance, are neglected.  However, it is
easy to see from our discussion that such details are not going to change
the essential features of our results, though they would determine
explicitly the specific ``large-$x$" region of the parameter space that
could be ruled out. The model has a rich spectrum of new particles at the 
$\mu_{\scriptscriptstyle D}$ or SCPV scale. Until such experimental data 
become available, a detailed study of the parameter space may not be feasible.

While the weak CP aspects of this model have been analyzed
in the non-supersymmetric setting as in Ref.~\cite{FN}, SUSY particles
could lead to new contributions through super-box and penguin diagrams. These 
contributions are in general subdominant, as are the Standard Model box diagrams.

Our model predicts a measureable neutron EDM, which could be close to the
present experimental bound for $x >.01$. The model also has
an extra pair of vector-like quark superfields, a new gauge boson, and a
number of neutral fermions and scalars with no direct couplings to the
Standard Model gauge bosons, all with masses around the TeV scale. This scale is
dictated by the weak CP phenomenology. Hence it offers a rich spectrum of new
particles to be discovered at the CERN Large Hadron Collider.
Because the $3\times 3$ Cabbibo-Kobayashi-Maskawa matrix is approximately real,
the model also predicts a flat unitarity triangle and the absence of
substantial CP violation in the $B$ system at future $B$ factories\cite{asponB}. 
Moreover, there will be a lack of any substantial CP violating effects in 
the up-quark sector.

\acknowledgements
The authors want to thank P.H. Frampton for being a constant source of support 
and encouragement. O.K. is in debt to colleagues in Rochester, where the major
part of this manuscript was finished. Special thanks go to M.~Bisset
for reading the manuscript. B.W. thanks P. Sikivie for discussions.
The authors were supported in part by the U.S. Department of
Energy under Grant DE-FG05-85ER-40219, Task B;
O.K. was also supported in part by the U.S. Department of
Energy under Grant DE-FG02-91ER-40685.\\

\newpage

\appendix
\section{Renormalization Group Equations}

Here we collect the modified one loop MSSM renormalization group equations
(RGE's)
to account for the extra vector-like chiral superfields and the new
Yukawa couplings in our model. In many cases we give only the extra
contributions and refer the interested reader to Ref.~\cite{MV} with
whom we share conventions. The complete two-loop renormalization group
equations for a softly broken supersymmetric theory can be found in
Refs.\cite{MV,RGE}.

The complete superpotential can be written as
\begin{eqnarray}
W & = & \bar{u}Y_u Q H_u + \bar{d} Y_d Q H_d + \bar{e} Y_e L H_d +
\mu_{{\scriptscriptstyle H}} H_u H_d + \bar{d}_i \gamma^{ia} D \chi_a
+ \mu_{{\scriptscriptstyle D}} \bar{D} D\nonumber\\
&&{} + \bar{\chi}_a \mu_{\chi}^{ab} \chi_{b}
+ \aleph \bar{\chi}_a \lambda^{ab} \chi_b
+ \lambda_{{\scriptscriptstyle \aleph}} \aleph^3
+ \mu_{{\scriptscriptstyle \aleph}} \aleph^2
\label{superpot}
\end{eqnarray}
where family indices are implicit except in the new Yukawa coupling
and $a,b = 1,2$.
The soft supersymmetry breaking Lagrangian can be written as
$\mathcal{L}_{\text{soft}} = \mathcal{L}_{\text{soft}}^
{\scriptscriptstyle{\text{MSSM}}} + \mathcal{L}_{\text{soft}}^{\text{extra}}$,
where
\begin{eqnarray}
- \mathcal{L}_{\text{soft}}^
{\scriptscriptstyle{\text{MSSM}}} & = & \hat{\bar{u}} h_u \hat{Q} H_u
+ \hat{\bar{d}} h_d \hat{Q} H_d + \hat{\bar{e}} h_e \hat{L} H_d
+ B_{{\scriptscriptstyle H}} \mu_{{\scriptscriptstyle H}}H_u H_d
+ \text{h.c.} \nonumber\\
&&{} + \hat{Q}^{\dagger}\tilde{m}_Q^2 \hat{Q}
+ \hat{L}^{\dagger}\tilde{m}_L^2 \hat{L}
+ \hat{\bar{u}} \tilde{m}_{\bar{u}}^2 \hat{\bar{u}}^{\dagger}
+ \hat{\bar{d}} \tilde{m}_{\bar{d}}^2 \hat{\bar{d}}^{\dagger}
+ \hat{\bar{e}} \tilde{m}_{\bar{e}}^2 \hat{\bar{e}}^{\dagger}\label{mssmsoft}\\
&&{} + m^2_{H_u} H_u^{\dagger} H^{}_u
+ m^2_{H_d} H_d^{\dagger} H^{}_d~, \nonumber
\end{eqnarray}
and
\begin{eqnarray}
- \mathcal{L}_{\text{soft}}^
{\text{extra}} & = & \hat{\bar{d}}_i h_{\gamma}^{ia} \hat{D} \chi_a
+ B_{{\scriptscriptstyle D}} \mu_{{\scriptscriptstyle D}}\hat{\bar{D}} \hat{D}
+ \bar{\chi} B_{\chi}^{ab} \chi
+ \bar{\chi}_a h_{\lambda}^{ab} \chi_b \aleph
+ h_{\scriptscriptstyle \aleph} \aleph^3
+ B_{\scriptscriptstyle \aleph} \aleph^2 + \text{h.c.}
\nonumber\\
&&{} + \hat{\bar{D}} \tilde{m}_{\bar{\scriptscriptstyle D}}^2
\hat{\bar{D}}^{\dagger}
+ \hat{D}^{\dagger}\tilde{m}_{\scriptscriptstyle D}^2 \hat{D}
+ \bar{\chi}^{\dagger}_a \tilde{m}_{\bar{\chi} ab}^2
\bar{\chi}_b^{}
+ \chi_a^{} \tilde{m}_{\chi ab}^2 \chi_b^{\dagger}
+ \aleph^{\dagger} \tilde{m}_{\scriptscriptstyle \aleph}^2 \aleph~.
\label{extrasoft}
\end{eqnarray}
Note that  $B_{\chi}$ and
$B_{{\scriptscriptstyle \aleph}}$ are defined in a different way from
$B_{{\scriptscriptstyle D}}$ and $B_{{\scriptscriptstyle H}}$;
the former have dimension (mass)$^2$ and are analogs of
$B_{{\scriptscriptstyle D}} \mu_{{\scriptscriptstyle D}}$
and $B_{{\scriptscriptstyle H}} \mu_{{\scriptscriptstyle H}}$.
Also, $B_{\chi}$ is a $2\times 2$ matrix.
Finally we have supersymmetry breaking gaugino masses
$M_a (a = 1,2,3)$ for the MSSM and a possible gaugino mass $M_A$
under ${\mathrm{U(1)}}_A$.

We give the MSSM one loop RGE's for these interactions below.
The gauge couplings are computed to two loops
%to investigate unification issues
and are given by
\begin{equation}
\frac{dg_a}{dt} = \frac{g_a^3}{16\pi^2} B^{(1)}_a + \frac{g_a^3}{(16\pi^2)^2}
\left(\sum_{b=1}^{3} B^{(2)}_{ab} g_b^2 - \sum_{x=u,d,e,\gamma} C^x_a
{\mathrm{Tr}}(Y_x^{\dagger} Y_x^{})\right)~,\label{gaugerge}
\end{equation}
where $B^{(1)} = (\frac{33}{5} + \frac{2}{5}N_D + \frac{3}{5}N_L,
1 + N_L, -3 + N_D)$,
\begin{equation}
B^{(2)} = \left(\begin{array}{ccc}
\frac{199}{25} + \frac{8}{75}N_D
+ \frac{9}{25}N_L & \frac{27}{5} + \frac{9}{5}N_L & \frac{88}{5}
+ \frac{32}{15}N_D \\
\frac{9}{5} + \frac{3}{5}N_L & 25 + 7 N_L & 24 \\
\frac{11}{5} + \frac{4}{15}N_D & 9 & 14 + \frac{34}{3}N_D \end{array} \right)~,
\end{equation}
and
\begin{equation}
C^{u,d,e,\gamma} = \left( \begin{array}{cccc}
\frac{26}{5} & \frac{14}{5} & \frac{18}{5} & \frac{4}{5}\\
6 & 6 & 2 & 0 \\
4 & 4 & 0 & 2
\end{array} \right)~.
\end{equation}
In the above we have allowed for the possibility of $N_D$ heavy
vector-like pairs of charge $\frac{1}{3}$ color triplets [SU(2) singlets]
and $N_L$ such pairs of hypercharge $-1$ SU(2) doublets (color singlets).
These include both the extra mirror pairs $D + \bar{D}$ and $L + \bar{L}$
which can interact directly with MSSM fields, but also extra mirror
pairs originating in the messenger sector at higher scales. We have
for simplicity omitted the possible Yukawa interactions of the extra mirror
lepton doublets and the two loop ${\mathrm{U(1)}}_A$ contributions.
In the actual computations, we use $N_D=1$, $N_L=0$.
At one loop we also have
\begin{equation}
16\pi^2\frac{dg_A}{dt} = 10 g_A^3~,\label{garge}
\end{equation}
when the ${\mathrm{U(1)}}_A$ is present.

Using the above definitions, the two loop gaugino mass equations are
\begin{eqnarray}
\frac{dM_a}{dt} & = & \frac{2g_a^2}{16\pi^2} B^{(1)}_a M_a
+ \frac{2g_a^3}{(16\pi^2)^2} \Big[ \sum_{b=1}^{3} B^{(2)}_{ab} g_b^2
\left(M_a + M_b\right) \nonumber\\
&&{} + \sum_{u,d,e,\gamma} C^x_a
\left({\mathrm{Tr}}(Y^{\dagger}_x h_x^{})
- M_a {\mathrm{Tr}}(Y^{\dagger}_x Y_x^{})
\right)\Big]~.\label{mgauginorge}
\end{eqnarray}
At one loop we have
\begin{equation}
16\pi^2\frac{dM_A}{dt} = 20 g_A^2 M_A~.\label{marge}
\end{equation}

The running down quark Yukawa matrix is modified by the presence of the
$D$-$\bar{d}$ couplings $\gamma^{ia}$ which are treated as $3\times 2$
matrices below.
We have for the up and down one loop Yukawas:
\begin{eqnarray}
16\pi^2 \frac{dY_d}{dt} & = & Y_d\left(
{\mathrm{Tr}}\left(3Y_d^{\dagger}Y_d^{} + Y_e^{\dagger}Y_e^{}\right) +
3Y_d^{\dagger}Y_d^{} + Y_u^{\dagger}Y_u^{} - \frac{16}{3}g_3^2 - 3 g_2^2
-\frac{7}{15}g_1^2\right) + \gamma\gamma^{\dagger}Y_d~,\label{ydrge}\\
16\pi^2 \frac{dY_u}{dt} & = & Y_u\left(
3{\mathrm{Tr}}\left(Y_u^{\dagger}Y_u^{}\right) +
3Y_u^{\dagger}Y_u^{} + Y_d^{\dagger}Y_d^{} - \frac{16}{3}g_3^2 - 3 g_2^2
-\frac{13}{15}g_1^2\right)~,\label{yurge}
\end{eqnarray}
and for $\gamma$:
\begin{equation}
16\pi^2\frac{d\gamma}{dt} = \gamma\left(
{\mathrm{Tr}}\left(\gamma^{\dagger}\gamma\right) +
2\gamma^{\dagger}\gamma + \lambda^{\dagger}\lambda
 - \frac{16}{3}g_3^2 -\frac{4}{15}g_1^2 -
4 g_A^2\right) + Y_d^{} Y_d^{\dagger}\gamma~,\label{gammarge}
\end{equation}
where for $\gamma$ we have included the effect of a possible extra
${\mathrm{U(1)}}_A$ as described in the text.

The running supersymmetric $\mu_{{\scriptscriptstyle D}}$ parameter is given by
\begin{equation}
16\pi^2\frac{d\mu_{{\scriptscriptstyle D}}}{dt} =
\mu_{{\scriptscriptstyle D}}\left({\mathrm{Tr}}\left(\gamma^{\dagger}
\gamma\right) - \frac{16}{3}g_3^2 - \frac{4}{15}g_1^2 -
4g_A^2\right)~.\label{mudrge}
\end{equation}
The equation for the corresponding soft mass parameter is
\begin{equation}
16\pi^2\frac{dB_{{\scriptscriptstyle D}}}{dt} =
2 {\mathrm{Tr}}\left(\gamma^{\dagger}
h_{\gamma}\right) + \frac{32}{3}g_3^2M_3 + \frac{8}{15}g_1^2M_1
+ 8g_A^2 M_A~.\label{bdrge}
\end{equation}
Similar parameters for the SCPV sector have
\begin{equation}
16\pi^2\frac{d\mu_{{\scriptscriptstyle \chi}}}{dt} =
\mu_{{\scriptscriptstyle \chi}}\lambda^{\dagger}\lambda
+ \lambda\lambda^{\dagger} \mu_{{\scriptscriptstyle \chi}}
+3 \mu_{{\scriptscriptstyle \chi}}\gamma^{\dagger}\gamma
-4g_A^2 \mu_{{\scriptscriptstyle \chi}}~,
\end{equation}
\begin{eqnarray}
16\pi^2\frac{dB_{{\scriptscriptstyle \chi}}}{dt} &=&
B_{{\scriptscriptstyle \chi}} \left(\lambda^{\dagger} \lambda
+ 3 \gamma^{\dagger} \gamma \right)
 + \lambda \lambda^{\dagger} B_{{\scriptscriptstyle \chi}}
 + \lambda\left[{\mathrm{Tr}}\left(\lambda^{\dagger}
 B_{{\scriptscriptstyle \chi}}\right) + \lambda_{\aleph}
 B_{{\scriptscriptstyle \aleph}}\right]
 \nonumber \\
& &
+ 2  \mu_{{\scriptscriptstyle \chi}}\lambda^{\dagger} h_{\lambda}
+ 2 h_{\lambda}\lambda^{\dagger} \mu_{{\scriptscriptstyle \chi}}
+ 6  \mu_{{\scriptscriptstyle \chi}}\gamma^{\dagger} h_{\gamma}
- 4\left( B_{{\scriptscriptstyle \chi}} - 2 \mu_{{\scriptscriptstyle \chi}}
M_A\right) g_A^2~,
\end{eqnarray}
and
\begin{equation}
16\pi^2\frac{d\mu_{{\scriptscriptstyle \aleph}}}{dt} =
\mu_{{\scriptscriptstyle \aleph}}
\left[\lambda_{{\scriptscriptstyle \aleph}}^2
+2 {\mathrm{Tr}}\left(\lambda^{\dagger}\lambda\right)\right]~,
\end{equation}
\begin{eqnarray}
16\pi^2\frac{dB_{{\scriptscriptstyle \aleph}}}{dt} &=&
B_{{\scriptscriptstyle \aleph}} \left[ \lambda_{{\scriptscriptstyle \aleph}}^2
+  2 {\mathrm{Tr}}\left(\lambda^{\dagger}\lambda\right)\right]
+ \lambda_{\scriptscriptstyle \aleph}
\left[\lambda_{\scriptscriptstyle \aleph} B_{{\scriptscriptstyle \aleph}}
+2 {\mathrm{Tr}}\left(\lambda^{\dagger}B_{{\scriptscriptstyle \chi}}\right)\right]
\nonumber \\
& &
+ 2  \mu_{{\scriptscriptstyle \aleph}} \left[\lambda_{{\scriptscriptstyle \aleph}}
h_{{\scriptscriptstyle \aleph}}
+2 {\mathrm{Tr}}\left(\lambda^{\dagger}h_{\lambda}\right)\right]~.
\end{eqnarray}
There are also RGE's for the extra Yukawa couplings:
\begin{eqnarray}
16\pi^2\frac{d\lambda}{dt} & = & \lambda\left(
{\mathrm{Tr}}\left(\lambda^{\dagger}\lambda\right)
 + \frac{1}{2} \lambda_{{\scriptscriptstyle \aleph}}^2
 + 2 \lambda^{\dagger}\lambda
+ 3 \gamma^{\dagger}\gamma
-4g_A^2 \right)~,  \\
16\pi^2\frac{d\lambda_{{\scriptscriptstyle \aleph}}}{dt} & = &
\lambda_{{\scriptscriptstyle \aleph}}
\left[\lambda_{{\scriptscriptstyle \aleph}}^2 +
2 {\mathrm{Tr}}\left(\lambda^{\dagger}\lambda\right)   \right]~.
\end{eqnarray}
Note that $\mu_{{\scriptscriptstyle \chi}}~, B_{{\scriptscriptstyle \chi}}~,
\lambda$ and $h_{\lambda}$ are all $2\times 2$ matrices.

The relevant soft supersymmetry breaking trilinear coupling RGE's
are given by
\begin{eqnarray}
16\pi^2\frac{dh_d}{dt} & = & h_d\left(
{\mathrm{Tr}}\left(3Y_d^{\dagger}Y_d^{} + Y_e^{\dagger}Y_e^{}\right) + 5
Y_d^{\dagger}Y_d^{} + Y_u^{\dagger}Y_u^{} - \frac{16}{3}g_3^2 - 3 g_2^2
-\frac{7}{15}g_1^2\right) + \gamma\gamma^{\dagger}h_d\nonumber\\
&&{} + Y_d\Bigl({\mathrm{Tr}}\left(6Y_d^{\dagger}h_d^{}
+ 2Y_e^{\dagger}h_e^{}\right) + 4Y_d^{\dagger}h_d^{}
+ 2Y_u^{\dagger}h_u^{}\nonumber\\
&&{} + \frac{32}{3}g_3^2M_3 + 6 g_2^2M_2
+ \frac{14}{15}g_1^2M_1\Bigr) +
2 h_{\gamma}\gamma^{\dagger}Y_d~,\label{hdrge}\\
16\pi^2\frac{dh_u}{dt} & = & h_u\left(
3{\mathrm{Tr}}\left(Y_u^{\dagger}Y_u^{}\right) + 5Y_u^{\dagger}Y_u^{}
+ Y_d^{\dagger}Y_d^{} - \frac{16}{3}g_3^2 - 3 g_2^2
-\frac{13}{15}g_1^2\right) \nonumber\\
&&{} + Y_u\Bigl(6{\mathrm{Tr}}\left(Y_u^{\dagger}h_u^{}\right)
+ 4Y_u^{\dagger}h_u^{}
+ 2Y_d^{\dagger}h_d^{}\nonumber\\
&&{} + \frac{32}{3}g_3^2M_3 + 6 g_2^2M_2
+ \frac{26}{15}g_1^2M_1\Bigr)~,\label{hurge}\\
16\pi^2\frac{dh_{\gamma}}{dt} & = & h_{\gamma}\left(
{\mathrm{Tr}}\left(\gamma^{\dagger}\gamma\right)
+ 3\gamma^{\dagger}\gamma + \lambda^{\dagger} \lambda
- \frac{16}{3}g_3^2
- \frac{4}{15}g_1^2 - 4 g_A^2\right)
+ Y_d^{} Y_d^{\dagger}h_{\gamma}\nonumber\\
&&{} + \gamma\left(2{\mathrm{Tr}}\left(\gamma^{\dagger}h_{\gamma}\right)
+ 3\gamma^{\dagger}h_{\gamma} + \lambda^{\dagger} h_{\lambda}
+ \frac{32}{3}g_3^2M_3
+ \frac{8}{15}g_1^2M_1 + 8 g_A^2 M_A\right) \nonumber \\
&&{} +2 h_d^{} Y_d^{\dagger}\gamma~, \label{hgammarge} \\
16\pi^2\frac{dh_{\lambda}}{dt} & = & h_{\lambda}\left(
{\mathrm{Tr}}\left(\lambda^{\dagger}\lambda\right)
+ 3\gamma^{\dagger}\gamma + 3 \lambda^{\dagger} \lambda
+ \frac{1}{2} \lambda_{{\scriptscriptstyle \aleph}}^2
- 4 g_A^2\right) \nonumber\\
&&{} + \lambda\left(2{\mathrm{Tr}}\left(\lambda^{\dagger}h_{\lambda}\right)
+ 6\gamma^{\dagger}h_{\gamma} + 3 \lambda^{\dagger} h_{\lambda}
+ \lambda_{{\scriptscriptstyle \aleph}} h_{{\scriptscriptstyle \aleph}}
 + 8 g_A^2 M_A\right) ~,\\
16\pi^2\frac{dh_{{\scriptscriptstyle \aleph}}}{dt} & = &
 h_{{\scriptscriptstyle \aleph}}\left(\frac{9}{2}
 \lambda_{{\scriptscriptstyle \aleph}}^2
 + 3 {\mathrm{Tr}}\left(\lambda^{\dagger}\lambda\right)\right)
 + 6  \lambda_{{\scriptscriptstyle \aleph}}
  {\mathrm{Tr}}\left(\lambda^{\dagger}h_{\lambda}\right)\; .
\end{eqnarray}

Finally we give the modifications of the MSSM soft hermitian quadratic
mass parameters (see Ref.\cite{MV} for a complete description).
All of the MSSM RGE's
have the following change in a D-term contribution: the factor $\mathcal{S}$
defined in Eq.(4.27) of \cite{MV} is now given by
\begin{equation}
{\mathcal{S}} = m_{H_u}^2 - m_{H_d}^2 -
\tilde{m}_D^2 + \tilde{m}_{\bar{D}}^2
+ {\mathrm{Tr}}\left(\tilde{m}_Q^2 - \tilde{m}_L^2 -
2\tilde{m}_{\bar{u}}^2 + \tilde{m}_{\bar{d}}^2 +
\tilde{m}_{\bar{e}}^2\right)~.\label{bigs}
\end{equation}
Besides this modification, the equation for $\tilde{m}_{\bar{d}}^2$ has the
only nontrivial change due to the coupling $\gamma$:
\begin{eqnarray}
16\pi^2\frac{d\tilde{m}_{\bar{d}}^2}{dt} & = &
\left(2\tilde{m}_{\bar{d}}^2 + 4m_{H_d}^2\right)Y_d^{} Y_d^{\dagger}
+ 4 Y_d^{} \tilde{m}_{Q}^2 Y_d^{\dagger}
+ 2 Y_d^{} Y_d^{\dagger}\tilde{m}_{\bar{d}}^2
+ \gamma\gamma^{\dagger}\tilde{m}_{\bar{d}}^2
+ \tilde{m}_{\bar{d}}^2\gamma\gamma^{\dagger}\nonumber\\
&&{} + 2 \tilde{m}_D^2 \gamma\gamma^{\dagger}
+ 2 \gamma\tilde{m}_{\chi}^2\gamma^{\dagger}
+ 4h_d^{} h_d^{\dagger} + 2h_{\gamma}^{}h_{\gamma}^{\dagger}\nonumber\\
&&{} - \frac{32}{3}g_3^2\left|M_3\right|^2
- \frac{8}{15}g_1^2\left|M_1\right|^2
+ \frac{2}{5}g_1^2{\mathcal{S}}~,\label{msqdbarrge}
\end{eqnarray}
where $\tilde{m}_{\chi}^2$ is a $2\times 2$ matrix.
The equations for the new soft squark masses are
\begin{eqnarray}
16\pi^2\frac{d\tilde{m}_{D}^2}{dt} & = & 2
{\mathrm{Tr}}\left(\tilde{m}_{D}^2\gamma^{\dagger}\gamma
+ \gamma\tilde{m}_{\chi}^2\gamma^{\dagger}
+ \gamma^{\dagger}\tilde{m}_{\bar{d}}^2\gamma
+ h_{\gamma}^{\dagger}h_{\gamma}^{}\right)\nonumber\\
&&{} - \frac{32}{3}g_3^2\left|M_3\right|^2
- \frac{8}{15}g_1^2\left|M_1\right|^2
- 8 g_A^2\left|M_A\right|^2 - \frac{2}{5}g_1^2{\mathcal{S}}
+ 2 g_A^2{\mathcal{S_A}}~,\label{msqbigdrge}\\
16\pi^2\frac{d\tilde{m}_{\bar{D}}^2}{dt} & = &
- \frac{32}{3}g_3^2\left|M_3\right|^2
- \frac{8}{15}g_1^2\left|M_1\right|^2
- 8 g_A^2\left|M_A\right|^2 + \frac{2}{5}g_1^2{\mathcal{S}}
- 2 g_A^2{\mathcal{S}}_A~,\label{msqbigdbarrge}\\
16\pi^2\frac{d\tilde{m}_{\chi}^2}{dt} & = &
3\gamma^{\dagger}\gamma \tilde{m}_{\chi}^2
+ 3\tilde{m}_{\chi}^2\gamma^{\dagger}\gamma
+ 6\gamma^{\dagger}\gamma \tilde{m}_{D}^2
+ 6\gamma^{\dagger}\tilde{m}_{\bar{d}}^2\gamma
+ 6 h_{\gamma}^{\dagger} h_{\gamma}^{}\label{msqchirge}\\
&&{} + \lambda^{\dagger}\lambda \tilde{m}_{\chi}^2
+ \tilde{m}_{\chi}^2\lambda^{\dagger}\lambda
+ 2\lambda^{\dagger}\lambda \tilde{m}_{{\scriptscriptstyle \aleph}}^2
+ 2\lambda^{\dagger}\tilde{m}_{\bar{\chi}}^2\lambda
+ 2 h_{\lambda}^{\dagger} h_{\lambda}^{}
- 8 g_A^2\left|M_A\right|^2
- 2 g_A^2{\mathcal{S}}_A~,\nonumber \\
16\pi^2\frac{d\tilde{m}_{\bar{\chi}}^2}{dt} & = &
\tilde{m}_{\bar{\chi}}^2 \lambda \lambda^{\dagger}
+ \lambda \lambda^{\dagger} \tilde{m}_{\bar{\chi}}^2
+ 2\lambda \lambda^{\dagger} \tilde{m}_{{\scriptscriptstyle \aleph}}^2
+ 2\lambda \tilde{m}_{\chi}^2 \lambda^{\dagger}
+ 2  h_{\lambda}^{}h_{\lambda}^{\dagger}
- 8 g_A^2\left|M_A\right|^2
+ 2 g_A^2{\mathcal{S}}_A~, \\
16\pi^2\frac{d\tilde{m}_{{\scriptscriptstyle \aleph}}^2}{dt} & = &
\left( 3\lambda_{{\scriptscriptstyle \aleph}}^2 +
 2 {\mathrm{Tr}}\lambda^{\dagger} \lambda \right)
 \tilde{m}_{{\scriptscriptstyle \aleph}}^2
+ 2 {\mathrm{Tr}}\left(\lambda \tilde{m}_{\chi}^2 \lambda^{\dagger}\right)
+ 2 {\mathrm{Tr}}\left(\lambda^{\dagger} \tilde{m}_{\bar{\chi}}^2 \lambda\right)
\nonumber \\
&&{}+ h_{{\scriptscriptstyle \aleph}}^2
+ 2 {\mathrm{Tr}} h_{\lambda}^{\dagger}h_{\lambda}^{}~,
\end{eqnarray}
where ${\mathcal{S}}_A = 3\tilde{m}_{D}^2 - 3\tilde{m}_{\bar{D}}^2
+ {\mathrm{Tr}}(\tilde{m}_{\bar{\chi}}^2 - \tilde{m}_{\chi}^2)$.

\section{Some numerical results}

We collect here some of the numerical results in our sample calculation. Recall
that we use $M_{mess}=50$ TeV, $\mu_{\scriptscriptstyle D}=
\mu_{\scriptscriptstyle \chi}=500$ GeV, and random values of $\gamma$'s
and $\lambda$'s in the ranges  $.005 - .01$ and $ .1 -.8$ respectively.
As inputs we also used the values $M_t = 175$ GeV and $\alpha_s = 0.12$.

The electroweak-symmetry breaking solution is obtained with
\begin{equation}
\tan\!\beta = 43.18\; , \quad \mu_H = -370.5{\rm\ GeV}\; ,
\quad B_H = 3.938 {\rm\ GeV}\; .
\end{equation}
In the following, we concentrate on the down-sector
as it is the only one of relevance to the understanding the $\bar{\theta}$
value.
Average values of the squared left- and right-handed squark masses
are $5.337\times 10^5$ GeV$^2$ and $5.031\times 10^5$ GeV$^2$, respectively,
and $\bar{A}_d$ is $-243.5$ GeV. The lack of
proportionality of the $A_d$-terms is given by
\begin{equation}
\delta\!A/\tilde{m}_{sq} =
\left( \begin{array}{ccc}
0. & 2.12\times 10^{-8} & 8.45\times 10^{-7}\\
 1.18\times 10^{-9} & 6.39\times 10^{-7} & 4.02\times 10^{-7}\\
-2.37\times 10^{-7} & 1.73\times 10^{-6} & 1.63\times 10^{-2}
\end{array}\right) \; ,
\end{equation}
while degeneracy violations are given by
\begin{equation}
\delta \tilde{m}_{d}^2/\tilde{m}_{sq}^2 =
\left( \begin{array}{ccc}
0. & 2.54\times 10^{-5} & -6.13\times 10^{-4}\\
 2.54\times 10^{-5} & -2.53\times 10^{-4} & 4.42\times 10^{-3}\\
-6.13\times 10^{-4} & 4.42\times 10^{-3} & -0.172
\end{array}\right) \; ,
\end{equation}
and
\begin{equation}
\delta \tilde{m}_{\bar{d}}^2/\tilde{m}_{sq}^2 =
\left( \begin{array}{ccc}
0. & -8.22\times 10^{-6} & -1.12\times 10^{-5}\\
 -8.22\times 10^{-6} & -1.59\times 10^{-4} & -1.17\times 10^{-5}\\
-1.12\times 10^{-5} & -1.17\times 10^{-5} & -0.135
\end{array}\right) \; ,
\end{equation}
where $\tilde{m}_{sq}^2$ is the average squark mass.
In addition, we have
\begin{eqnarray}
\delta \tilde{m}_{\bar{D}}^2/\tilde{m}_{sq}^2 &=& 1.65\times 10^{-5}\; , \\
\delta \tilde{m}_D^2/\tilde{m}_{sq}^2 &=& -5.64\times 10^{-2}\; .
\end{eqnarray}
The proportional part of the $A$-terms for the $\gamma$ Yukawas is
very close to the $B_D$ value, given by
\begin{equation}
\bar{A}_{\gamma} \sim B_D = -248.1\, {\rm GeV}\; ,
\end{equation}
with a difference of only $4.27\times 10^{-2}$ GeV.
Some other quantities of interest are :
\[
M_g (\equiv M_3) = 478.1 {\rm\ GeV}\; , \quad
x = 1.213\times 10^{-2} \; ;
\]
and, as defined by Eq.(\ref{dagam}),
\begin{equation}
\delta\! A_{\gamma} = 4.782\times 10^{-4} {\rm GeV}\; ,
\quad {\rm Im}(\mbox{\boldmath $a^{\dag}c$})
= 5.531\times 10^{-3} \; .
\end{equation}

\clearpage

{\bf Table Captions.}\\

Table 1: Analysis of the major 1-loop $\bar{\theta}$ contributions and
numerical results from the sample run. Table 1a contains contributions
from gluino mass corrections; Table 1b contains those from quark mass
corrections. Entry 1 of Table~1a and entries 9, and 10 of
Table~1b are shown together with
explicit splittings of  $M_5^2$ according to Eq.(3.6) below the first
lines. Numerical results given in the last column do not include the
$\left\langle F_{\chi}\right\rangle$ term contributions, but otherwise are complete,
{\it i.e.} they include all other numerical factors from color indices
summation, momentum loop integrals, and full summation over
family indices ($i,j,k$) so that the full $\bar{\theta}$ value
without the $F$-term contributions, apart from
some unlisted subdominating terms, is given by the sum of
all the entries.

Table 2:  Estimates of the $\left\langle F_{\chi}\right\rangle$ term and its contribution
to $\bar{\theta}$, for our sample run and a few runs with different
$\gamma$ and $\lambda$ inputs ($\mu_{\scriptscriptstyle D}$ and $\mu_{\chi}$'s
are all set at $500$ GeV, $M_{mess}$ at $50$ TeV).
Note that the entries $B_{\chi}$,
$\tilde{m}^2_{\bar{\chi}}$, $h_{\lambda} \left\langle \chi\right\rangle / \lambda$,
and $h_{\gamma} \left\langle \chi\right\rangle  \gamma / 16\pi^2$ are our  $\left\langle F_{\chi}\right\rangle$
estimates, as discussed;
all these are quantities of dimension
(mass)$^2$ in units of GeV$^2$ (not shown explicitly). The 
$\left\langle F_{\chi}\right\rangle$ estimates and its contributions to $\bar{\theta}$ are meant
to be upper bounds.
Overall  $\bar{\theta}$ contributions
from gluino and quark mass corrections without the $F$-term are
also listed.

\bigskip
\bigskip

{\bf Figure Captions.}\\

Fig.1 : 1-loop mass-correction diagrams leading to $\bar{\theta}$ contributions.
(a) 1-loop quark mass; (b) 1-loop gluino mass.

Fig.2 :  Diagrams giving estimates of $\left\langle F_{\chi} \right\rangle$ magnitudes.
Note that
$\left\langle \chi \right\rangle$, $\left\langle \bar{\chi} \right\rangle$, $\mu_{\chi}$ and
all propagator masses in the diagrams can be taken as around the
same scale, namely $\mu_{\scriptscriptstyle D}$; a SUSY breaking
vertex or mass insertion is required in each case, as shown.

%\onecolumn

%\clearpage

\begin{figure}

\vspace{6in}
\includegraphics{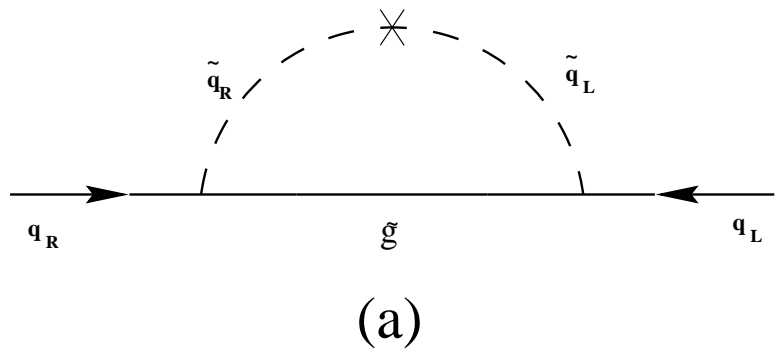}
%\vspace{4in}
\includegraphics{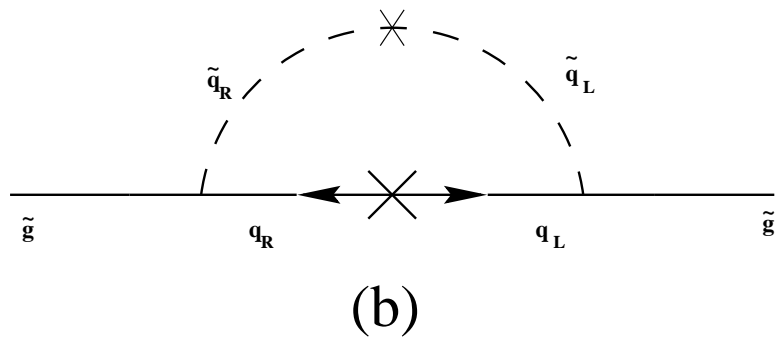}
\caption{1-loop mass-correction diagrams leading to $\bar{\theta}$ contributions.
(a) 1-loop quark mass; (b) 1-loop gluino mass.}
\end{figure}

\clearpage

\begin{figure}

\vspace{10in}
\includegraphics{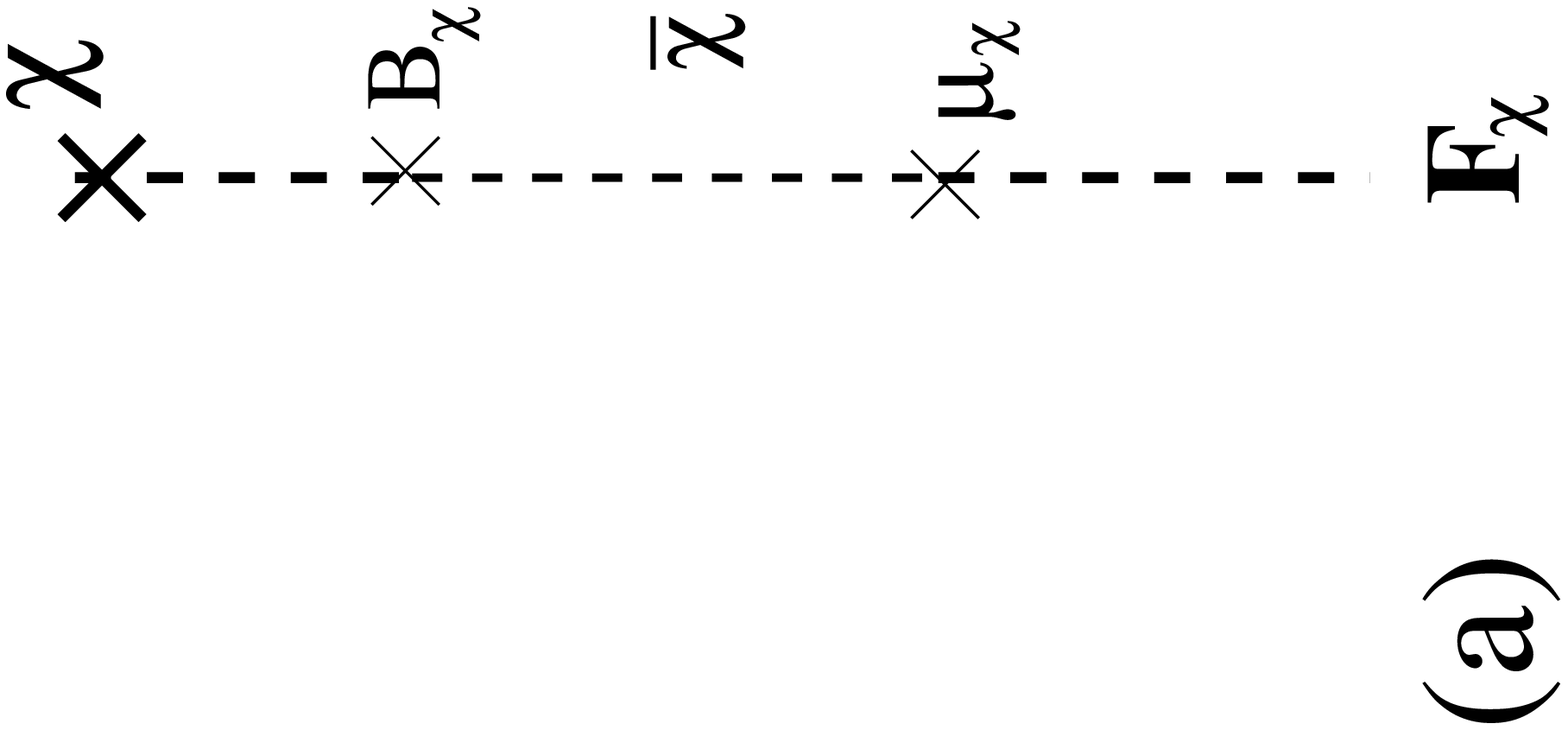}
\includegraphics{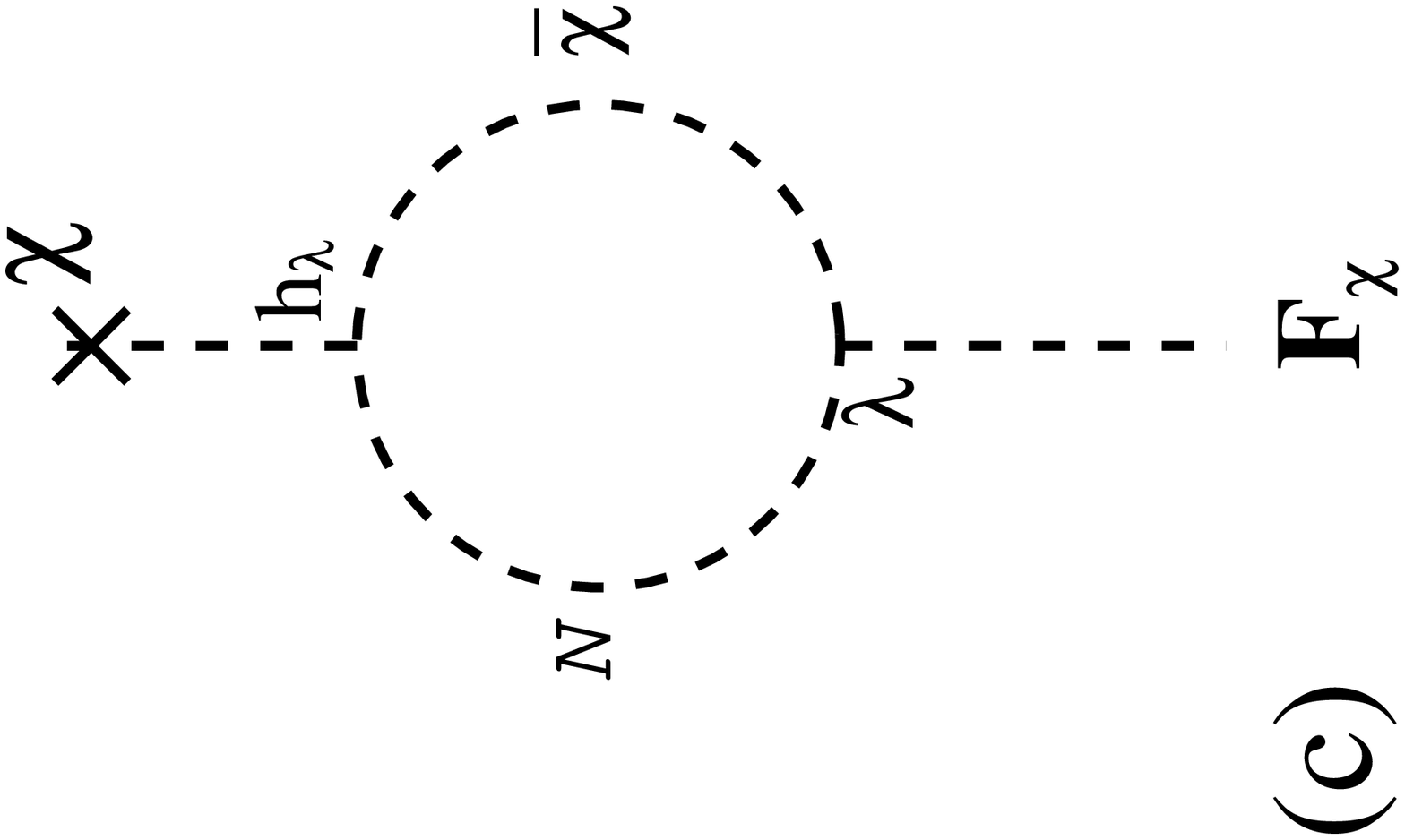}
\includegraphics{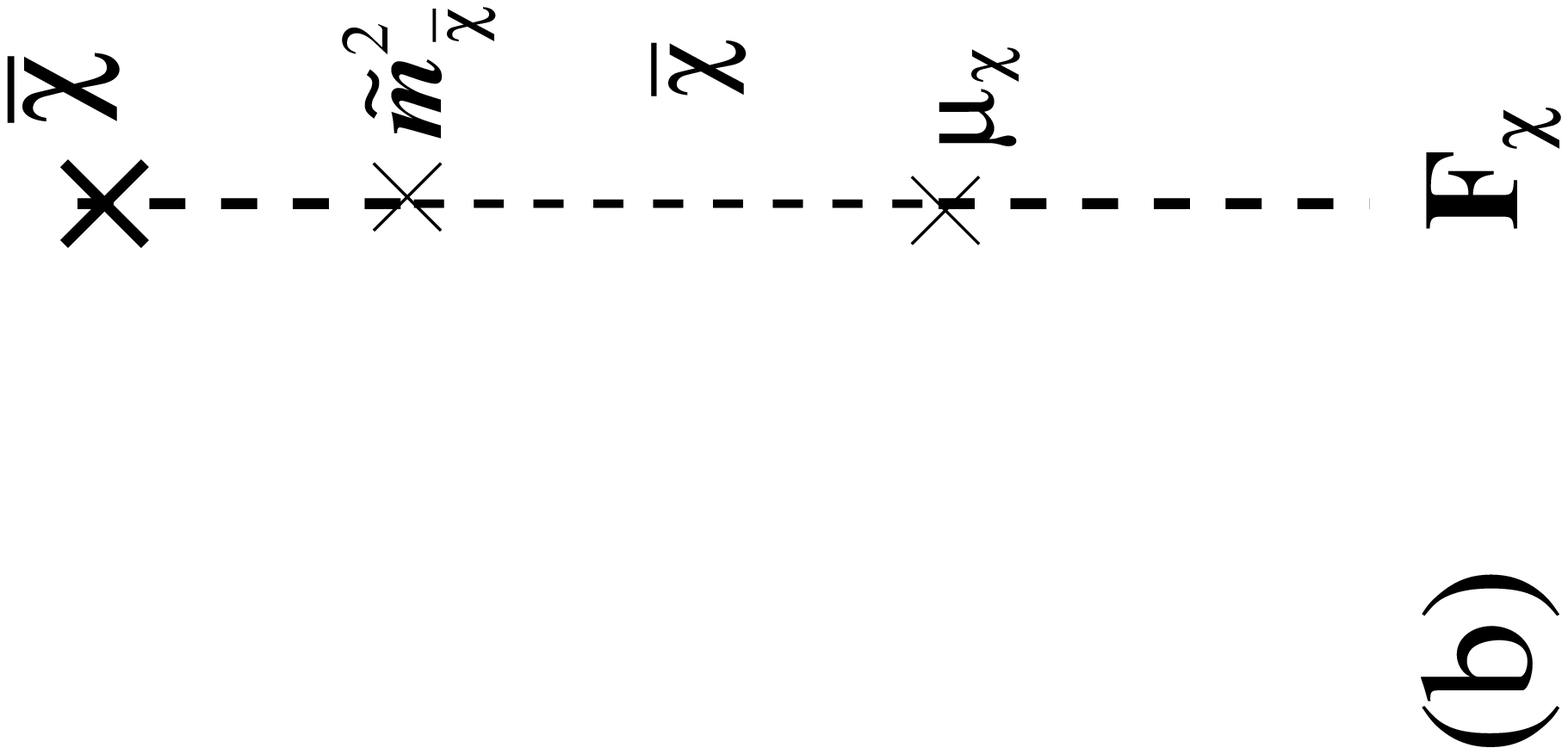}
\includegraphics{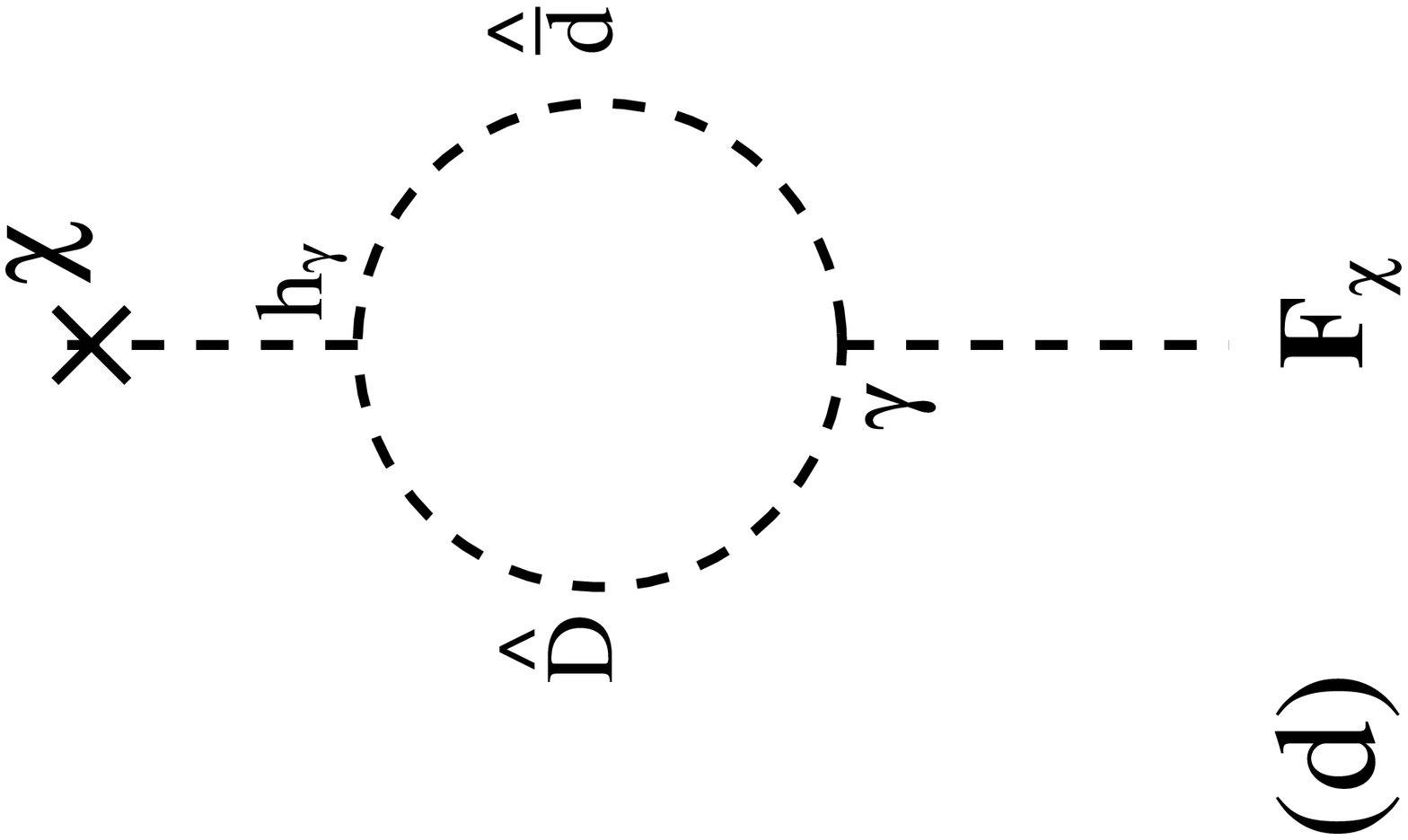}
\caption{Diagrammatic estimates of $\left\langle F_{\chi} \right\rangle$ magnitudes. Note that
$\left\langle \chi_a \right\rangle$ and all the mass insertions in the diagrams are around the
same scale, namely $\mu_{\scriptscriptstyle D}$; a SUSY breaking
vertex or mass insertion is required in each case, as shown.  }
\end{figure}

\clearpage

\small

\noindent
Table 1: Analysis of the major 1-loop $\bar{\theta}$ contributions and
numerical results from the sample run. Table 1a contains contributions
from gluino mass corrections; Table 1b contains those from quark mass
corrections. Entry 1 of Table~1a and entries 9, and 10 of
Table~1b are shown together with
explicit splittings of  $M_5^2$ according to Eq.(3.6) below the first
lines. Numerical results given in the last column do not include the
$\left\langle F_{\chi}\right\rangle$ term contributions, but otherwise are complete,
{\it i.e.} they include all other numerical factors from color indices
summation, momentum loop integrals, and full summation over
family indices ($i,j,k$) so that the full $\bar{\theta}$ value
without the $F$-term contributions, apart from
some unlisted subdominating terms, is given by the sum of
all the entries. \\[.3in]
\noindent
Table 1a: Gluino mass correction contributions\\[.2in]
\begin{tabular}{c||cccccccccc||c}\hline\hline
No. & \multicolumn{10}{c||}{ factors  of $\bar{\theta}$ contribution} &
 \\ \hline
%g1
(1) & $\frac{\alpha_s}{4\pi} x$  & $\frac{\mu_{\scriptscriptstyle D}}{\tilde{m}_{sq}}$ & &
$\frac{\tilde{m}_{sq}}{M_g}$  & &  & & &
$\frac{M_5^2}{\tilde{m}_{sq}^2}$   &  ${\rm Im}(a^{*}_i b^i)$  &
 $-5.17\times 10^{-15}$\\
 --& $\frac{\alpha_s}{4\pi} x^2$  &  $\frac{\mu_{\scriptscriptstyle D}^2}{\tilde{m}_{sq}^2}$ & &
$\frac{\tilde{m}_{sq}}{M_g}$  & &  & & &
$\frac{\delta\!A_{\gamma}}{\tilde{m}_{sq}}$ &   ${\rm Im}(a^{*}_i c^i)$  &
- -\\
 --& $\frac{\alpha_s}{4\pi} x$  &  $\frac{\mu_{\scriptscriptstyle D}}{\tilde{m}_{sq}}$ & &
$\frac{\tilde{m}_{sq}}{M_g}$  & &  & & &
\multicolumn{2}{c||}{
$\frac{\gamma^{ia} |F_{\chi_a}|}{\tilde{m}_{sq}^2}$} &
(cf. table 2) \\
 %g7
%(2) & $\frac{\alpha_s}{4\pi} x$ &
%$\frac{\mu_{\scriptscriptstyle D}^3}{\tilde{m}_{sq}^3}$ &
%$\frac{B_{\scriptscriptstyle D}^2}{\tilde{m}_{sq}^2}$  &
%$\frac{\tilde{m}_{sq}}{M_g}$ & & & & &
%$\frac{M_5^2}{\tilde{m}_{sq}^2}$  &   ${\rm Im}(a^{*}_i b^i)$ &
%$< (1)$\\
%g7
(2) & $\frac{\alpha_s}{4\pi} x$ & $\frac{\mu_{\scriptscriptstyle D}}{\tilde{m}_{sq}}$ &
$\frac{\tilde{m}_{\scriptscriptstyle A}^2 m_i^2}{\tilde{m}_{sq}^4}$  &
$\frac{\tilde{m}_{sq}}{M_g}$ & & & & &
$\frac{M_5^2}{\tilde{m}_{sq}^2}$  &   ${\rm Im}(a^{*}_i b^i)$ &
$-1.53\times 10^{-14}$\\
%g4
%(3)  & $\frac{\alpha_s}{4\pi} x$  & $\frac{\mu_{\scriptscriptstyle D}}{\tilde{m}_{sq}}$  & &
%$\frac{\tilde{m}_{sq}}{M_g}$  & &  &
%$\frac{\delta\!  \tilde{m}_{\scriptscriptstyle D}^2}{\tilde{m}_{sq}^2}$    & &
%$\frac{M_5^2}{\tilde{m}_{sq}^2}$   &  ${\rm Im}(a^{*}_i b^i)$  &
%$-7.84\times 10^{-17}$\\
%g2
(3) & $\frac{\alpha_s}{4\pi} x^2$ & $\frac{\mu_{\scriptscriptstyle D}^2}{\tilde{m}_{sq}^2}$ &
$\frac{v_d  m_i}{\tilde{m}_{sq}^2}$  &
$\frac{\tilde{m}_{sq}}{M_g}$ & & & &
$\frac{\delta\! A^{ji}}{\tilde{m}_{sq}}$ & &   ${\rm Im}(a^{*}_j a^i)$ &
$1.53\times  10^{-18}$\\
%g10
(4) & $\frac{\alpha_s}{4\pi} x^2$ & $\frac{\mu_{\scriptscriptstyle D}^2}{\tilde{m}_{sq}^2}$ &
$\frac{\tilde{m}_{\scriptscriptstyle A} m_i^2}{\tilde{m}_{sq}^3}$  &
$\frac{\tilde{m}_{sq}}{M_g}$ & &
$\frac{\delta\!  \tilde{m}_{\bar{d}}^{2jk}}{\tilde{m}_{sq}^2}$ & &
 & &   ${\rm Im}(a^{*}_j a^i)$ &
$1.09\times  10^{-16}$ \\
%g3
(5)  & $\frac{\alpha_s}{4\pi} x$  &  $\frac{\mu_{\scriptscriptstyle D}}{\tilde{m}_{sq}}$ & &
$\frac{\tilde{m}_{sq}}{M_g}$  & &
$\frac{\delta\!  \tilde{m}_{\bar{d}}^{2ij}}{\tilde{m}_{sq}^2}$   &  & &
$\frac{M_5^2}{\tilde{m}_{sq}^2}$   &  ${\rm Im}(a^{*}_j b^i)$  &
$1.27\times  10^{-14}$\\
%g5
(6) & $\frac{\alpha_s}{4\pi} x^2$ & $\frac{\mu_{\scriptscriptstyle D}^2}{\tilde{m}_{sq}^2}$ &
$\frac{v_d  m_i}{\tilde{m}_{sq}^2}$  &
$\frac{\tilde{m}_{sq}}{M_g}$ & &
$\frac{\delta\!  \tilde{m}_{\bar{d}}^{2jk}}{\tilde{m}_{sq}^2}$ & &
$\frac{\delta\! A^{ki}}{\tilde{m}_{sq}}$ & &   ${\rm Im}(a^{*}_j a^i)$ &
$2.39\times  10^{-19}$ \\
%g6
(7) & $\frac{\alpha_s}{4\pi} x^2$ & $\frac{\mu_{\scriptscriptstyle D}^2}{\tilde{m}_{sq}^2}$ &
$\frac{v_d  m_i}{\tilde{m}_{sq}^2}$  &
$\frac{\tilde{m}_{sq}}{M_g}$ & & &
$\frac{\delta\!  \tilde{m}_{d}^{2ki}}{\tilde{m}_{sq}^2}$ &
$\frac{\delta\! A^{jk}}{\tilde{m}_{sq}}$ & &   ${\rm Im}(a^{*}_j a^i)$ &
$-5.99\times  10^{-18}$\\
%g7
(8) & $\frac{\alpha_s}{4\pi}$ &  &
$\frac{v_d  m_i}{\tilde{m}_{sq}^2}$  &
$\frac{\tilde{m}_{sq}}{M_g}$ & & & &
$\frac{\delta\! A^{ji}}{\tilde{m}_{sq}}$ &
$\frac{(M_5^2)^2}{\tilde{m}_{sq}^4}$ &   ${\rm Im}(b^{*}_j b^i)$ &
$8.43\times  10^{-20}$ \\
%g7
(9) & $\frac{\alpha_s}{4\pi} x$ & $\frac{\mu_{\scriptscriptstyle D}}{\tilde{m}_{sq}}$ &
$\frac{v_d  m_i \tilde{m}_{\scriptscriptstyle A}}{\tilde{m}_{sq}^3}$  &
$\frac{\tilde{m}_{sq}}{M_g}$ & & & &
$\frac{\delta\! A^{ji}}{\tilde{m}_{sq}}$ &
$\frac{M_5^2}{\tilde{m}_{sq}^2}$ &   ${\rm Im}(a^{*}_j a^i)$ &
$1.14\times  10^{-18}$ \\
%g8
(10) & $\frac{\alpha_s}{4\pi} x$ & $\frac{\mu_{\scriptscriptstyle D}}{\tilde{m}_{sq}}$ &
$\frac{v_d^2}{\tilde{m}_{sq}^2}$  &
$\frac{\tilde{m}_{sq}}{M_g}$ & & & &
$\frac{\delta\! A^{jk}}{\tilde{m}_{sq}} \frac{\delta\! A^{ik}}{\tilde{m}_{sq}}$ &
$\frac{M_5^2}{\tilde{m}_{sq}^2}$  &   ${\rm Im}(a^{*}_j b^i)$ &
$5.26\times  10^{-22}$\\
\hline\hline
\end{tabular}

\clearpage

\noindent
Table 1b: Quark mass correction contributions\\[.2in]
\begin{tabular}{c||cccccccccc||c}\hline\hline
No. & \multicolumn{10}{c||}{ factors  of $\bar{\theta}$ contribution} &
 \\ \hline
%q1
(1) & $\frac{\alpha_s}{4\pi} x$  & $\frac{\mu_{\scriptscriptstyle D}}{\tilde{m}_{sq}}$ & &
$\frac{M_g}{\tilde{m}_{sq}}$  & &  & & &
$\frac{M_5^2}{\tilde{m}_{sq}^2}$   &  ${\rm Im}(a^{*}_i b^i)$  &
$-1.52\times  10^{-15}$ \\
%q8
(2)  & $\frac{\alpha_s}{4\pi} x$  &  $\frac{\mu_{\scriptscriptstyle D}}{\tilde{m}_{sq}}$ &
$\frac{\tilde{m}_{\scriptscriptstyle A} B_{\scriptscriptstyle D}}{\tilde{m}_{sq}^2}$ &
$\frac{M_g}{\tilde{m}_{sq}}$  & &
   &  & &
$\frac{M_5^2}{\tilde{m}_{sq}^2}$   &  ${\rm Im}(a^{*}_i b^i)$  &
 $3.45\times  10^{-15}$ \\
%q2
(3) & $\frac{\alpha_s}{4\pi} x^2$ & $\frac{\mu_{\scriptscriptstyle D}^2}{\tilde{m}_{sq}^2}$ &
$\frac{v_d  m_i}{\tilde{m}_{sq}^2}$  &
$\frac{M_g}{\tilde{m}_{sq}}$ & & & &
$\frac{\delta\! A^{ji}}{\tilde{m}_{sq}}$ & &   ${\rm Im}(a^{*}_j a^i)$ &
$3.10\times  10^{-19}$ \\
%q3
(4)  & $\frac{\alpha_s}{4\pi} x$  &  $\frac{\mu_{\scriptscriptstyle D}}{\tilde{m}_{sq}}$ & $\frac{m_i}{m_j}$ &
$\frac{M_g}{\tilde{m}_{sq}}$  & &
$\frac{\delta\!  \tilde{m}_{\bar{d}}^{2ij}}{\tilde{m}_{sq}^2}$   &  & &
$\frac{M_5^2}{\tilde{m}_{sq}^2}$   &  ${\rm Im}(a^{*}_i b^j)$  &
$2.02\times  10^{-14}$ \\
%q6
%(5$'$)  & $\frac{\alpha_s}{4\pi} x$  &  $\frac{\mu_{\scriptscriptstyle D}}{\tilde{m}_{sq}}$ & &
%$\frac{M_g}{\tilde{m}_{sq}}$  & &
%$\frac{\delta\!  \tilde{m}_{\bar{\scriptscriptstyle D}}^2}{\tilde{m}_{sq}^2}$ & &  &
%$\frac{M_5^2}{\tilde{m}_{sq}^2}$   &  ${\rm Im}(a^{*}_i b^i)$  &
%$-3.47\times  10^{-17}$ \\
%q4
(5)  & $\frac{\alpha_s}{4\pi} x^2$  & $\frac{\mu_{\scriptscriptstyle D}^2}{\tilde{m}_{sq}^2}$  &
$\frac{v_d}{m_k}$ & $\frac{M_g}{\tilde{m}_{sq}}$  & &
$\frac{\delta\!  \tilde{m}_{\bar{d}}^{2jk}}{\tilde{m}_{sq}^2}$ & &
$\frac{\delta\! A^{ik}}{\tilde{m}_{sq}}$ & &   ${\rm Im}(a^{*}_i a^j)$ &
$2.00\times  10^{-15}$\\
%q4
(6)  & $\frac{\alpha_s}{4\pi} x^2$  & $\frac{\mu_{\scriptscriptstyle D}^2}{\tilde{m}_{sq}^2}$  &
$\frac{v_d}{m_j}$ & $\frac{M_g}{\tilde{m}_{sq}}$  & &
$\frac{\delta\!  \tilde{m}_{\bar{\scriptscriptstyle D}}^2}{\tilde{m}_{sq}^2}$ & &
$\frac{\delta\! A^{ij}}{\tilde{m}_{sq}}$ & &   ${\rm Im}(a^{*}_i a^j)$ &
$8.00\times  10^{-18}$ \\
%q16
(7)  & $\frac{\alpha_s}{4\pi} x^2$  & $\frac{\mu_{\scriptscriptstyle D}^2}{\tilde{m}_{sq}^2}$  &
$\frac{m_i \tilde{m}_{\scriptscriptstyle A}}{m_k \tilde{m}_{sq}}$   &
$\frac{M_g}{\tilde{m}_{sq}}$  & &
 $\frac{\delta\!  \tilde{m}_{\bar{d}}^{2kj}}{\tilde{m}_{sq}^2}$
 & $\frac{\delta\!  \tilde{m}_{d}^{2ik}}{\tilde{m}_{sq}^2}$ &
 & &   ${\rm Im}(a^{*}_i a^j)$ &
$-7.49\times  10^{-13}$ \\
%q17
(8)  & $\frac{\alpha_s}{4\pi} x^2$  & $\frac{\mu_{\scriptscriptstyle D}^2}{\tilde{m}_{sq}^2}$  &
$\frac{m_j \tilde{m}_{\scriptscriptstyle A}}{m_i \tilde{m}_{sq}}$ & $\frac{M_g}{\tilde{m}_{sq}}$  & &
$\frac{\delta\!  \tilde{m}_{\bar{\scriptscriptstyle D}}^2}{\tilde{m}_{sq}^2}$ & $\frac{\delta\!  \tilde{m}_{d}^{2ji}}{\tilde{m}_{sq}^2}$ &
 & &   ${\rm Im}(a^{*}_i a^j)$ &
$2.26\times  10^{-13}$ \\
%q5
(9)  & $\frac{\alpha_s}{4\pi} x$  &  $\frac{\mu_{\scriptscriptstyle D}}{\tilde{m}_{sq}}$ & $\frac{m_j}{m_i}$ &
$\frac{M_g}{\tilde{m}_{sq}}$  & &  &
$\frac{\delta\!  \tilde{m}_{d}^{2ji}}{\tilde{m}_{sq}^2}$ &  &
\multicolumn{2}{c||}{
${\rm Im}(a_j^*(\frac{M_5^2}{{\tilde{m}_{sq}^2}} b^i
- x \mu_{\scriptscriptstyle D}\frac{B_{\scriptscriptstyle D} }{\tilde{m}_{sq}^2} a^i))$} &
 $-5.74\times  10^{-15}$\\
-- & $\frac{\alpha_s}{4\pi} x^2$  &  $\frac{\mu_{\scriptscriptstyle D}^2}{\tilde{m}_{sq}^2}$ & $\frac{m_j}{m_i}$ &
$\frac{M_g}{\tilde{m}_{sq}}$  & &  &
$\frac{\delta\!  \tilde{m}_{d}^{2ji}}{\tilde{m}_{sq}^2}$ &  &
$\frac{\bar{A}_\gamma - B_{\scriptscriptstyle D} }{\tilde{m}_{sq}}$ &
${\rm Im}(a^{*}_j a^i)$ &
- -\\
 --& $\frac{\alpha_s}{4\pi} x^2$  &  $\frac{\mu_{\scriptscriptstyle D}^2}{\tilde{m}_{sq}^2}$ & $\frac{m_j}{m_i}$ &
$\frac{M_g}{\tilde{m}_{sq}}$  & &  &
$\frac{\delta\!  \tilde{m}_{d}^{2ji}}{\tilde{m}_{sq}^2}$ &  &
$\frac{\delta\! A_\gamma}{\tilde{m}_{sq}}$ &
${\rm Im}(a^{*}_j c^i)$ &
- - \\
  -- & $\frac{\alpha_s}{4\pi} x$  &  $\frac{\mu_{\scriptscriptstyle D}}{\tilde{m}_{sq}}$ & $\frac{m_j}{m_i}$ &
$\frac{M_g}{\tilde{m}_{sq}}$  & &  &
$\frac{\delta\!  \tilde{m}_{d}^{2ji}}{\tilde{m}_{sq}^2}$ &  &
\multicolumn{2}{c||}{$\frac{\gamma^{ia} |F_{\chi_a}|}{\tilde{m}_{sq}^2}$} &
(cf. table 2)\\
%q7
(10)  & $\frac{\alpha_s}{4\pi} $  &   & $\frac{v_d}{m_i}$ &
$\frac{M_g}{\tilde{m}_{sq}}$  & &  &  &
$\frac{\delta\! A^{ji}}{\tilde{m}_{sq}}$ &
$\frac{(M_5^2)^2}{\tilde{m}_{sq}^4}$ &   ${\rm Im}(b^{*}_j b^i)$ &
$1.78\times  10^{-16}$ \\
 --& $\frac{\alpha_s}{4\pi} x^2$  &  $\frac{\mu_{\scriptscriptstyle D}^2}{\tilde{m}_{sq}^2}$
 & $\frac{v_d \bar{A}_{\gamma}^2}{m_i \tilde{m}_{sq}^2}$ &
$\frac{M_g}{\tilde{m}_{sq}}$  & &  &  &
$\frac{\delta\! A^{ji}}{\tilde{m}_{sq}}$ &
 &  ${\rm Im}(a^{*}_j a^i)$ &
- -\\
-- & $\frac{\alpha_s}{4\pi} x^2$  &  $\frac{\mu_{\scriptscriptstyle D}^2}{\tilde{m}_{sq}^2}$
 & $\frac{v_d}{m_i}$ &
$\frac{M_g}{\tilde{m}_{sq}}$  & &  &  &
$\frac{\delta\! A^{ji}}{\tilde{m}_{sq}}$ & $\frac{\delta\! A_{\gamma}^2}{\tilde{m}_{sq}^2}$
 &  ${\rm Im}(c^{*}_j c^i)$ &
- -\\
--& $\frac{\alpha_s}{4\pi} x^2$  &  $\frac{\mu_{\scriptscriptstyle D}^2}{\tilde{m}_{sq}^2}$
 & $\frac{v_d \bar{A}_{\gamma}}{m_i \tilde{m}_{sq}}$ &
$\frac{M_g}{\tilde{m}_{sq}}$  & &  &  & 
$\frac{\delta\! A^{ji}}{\tilde{m}_{sq}}$ &
$\frac{\delta\! A_\gamma}{\tilde{m}_{sq}}$ &  ${\rm Im}(c^{*}_j a^i+a^{*}_j c^i)$ &
- -\\
-- & $\frac{\alpha_s}{4\pi} x$  &  $\frac{\mu_{\scriptscriptstyle D}}{\tilde{m}_{sq}}$
  & $\frac{v_d \bar{A}_{\gamma}}{m_i \tilde{m}_{sq}}$ &
$\frac{M_g}{\tilde{m}_{sq}}$  & &  &  & 
$\frac{\delta\! A^{ji}}{\tilde{m}_{sq}}$ & 
\multicolumn{2}{c||}{
$\frac{\gamma^{ia} |F_{\chi_a}|}{\tilde{m}_{sq}^2}$} &
(cf. table 2)\\
-- & $\frac{\alpha_s}{4\pi} x$  &  $\frac{\mu_{\scriptscriptstyle D}}{\tilde{m}_{sq}}$
  & $\frac{v_d}{m_i}$ &
$\frac{M_g}{\tilde{m}_{sq}}$  & &  &  & 
$\frac{\delta\! A^{ji}}{\tilde{m}_{sq}}$ & 
$\frac{\delta\! A_\gamma}{\tilde{m}_{sq}}$  &
$\frac{\gamma^{ia} |F_{\chi_a}|}{\tilde{m}_{sq}^2}$ &
(cf. table 2)\\
 -- & $\frac{\alpha_s}{4\pi} $  &   & $\frac{v_d}{m_i}$ &
$\frac{M_g}{\tilde{m}_{sq}}$  & &  &  &
$\frac{\delta\! A^{ji}}{\tilde{m}_{sq}}$ &
\multicolumn{2}{c||}{$\frac{\gamma^{ja} |F_{\chi_a}|}{\tilde{m}_{sq}^2}$
$\frac{\gamma^{ia} |F_{\chi_a}|}{\tilde{m}_{sq}^2}$} &
(cf. table 2)\\
%q7
(11) & $\frac{\alpha_s}{4\pi} x $  &  $\frac{\mu_{\scriptscriptstyle D}}{\tilde{m}_{sq}}$
 & $\frac{v_d B_{\scriptscriptstyle D}}{m_i \tilde{m}_{sq}}$ &
$\frac{M_g}{\tilde{m}_{sq}}$  & &  &  &
$\frac{\delta\! A^{ji}}{\tilde{m}_{sq}}$ &
$\frac{M_5^2}{\tilde{m}_{sq}^2}$ &   ${\rm Im}(b^{*}_j a^i)$ &
 $3.08\times  10^{-21}$\\
\hline\hline
\end{tabular}

\clearpage
\normalsize
%\small

\noindent
Table 2:  Estimates of the $\left\langle F_{\chi}\right\rangle$ term and its contribution
to $\bar{\theta}$, for our sample run and a few runs with different
$\gamma$ and $\lambda$ inputs ($\mu_{\scriptscriptstyle D}$ and $\mu_{\chi}$'s
are all set at $500$ GeV, $M_{mess}$ at $50$ TeV).
Note that the entries $B_{\chi}$,
$\tilde{m}^2_{\bar{\chi}}$, $h_{\lambda} \left\langle \chi\right\rangle / \lambda$,
and $h_{\gamma} \left\langle \chi\right\rangle  \gamma / 16\pi^2$ are our  $\left\langle F_{\chi}\right\rangle$
estimates, as discussed;
all these are quantities of dimension
(mass)$^2$ in units of GeV$^2$ (not shown explicitly). The 
$\left\langle F_{\chi}\right\rangle$ estimates and its contributions to $\bar{\theta}$ are meant
to be upper bounds.
Overall  $\bar{\theta}$ contributions
from gluino and quark mass corrections without the $F$-term are
also listed. \\[.3in]
\begin{tabular}{c|cccccc}\hline\hline
%No. & 	     			4 & 		8 & 		10 	&	11	&	17	& 	19\\ \hline
No. & 	     			1 & 	2 (sample) & 		3	&	4	&	5	& 	6\\ \hline
$x$ 			& $.0081$	& $.012$	& $.02$		& $.068$	& $.0086$ 	&  $.0077$ \\
$\lambda$		& $ .23-.44$	& $ .18-.68$	& $ .53-.77$ 	& $ .27-.61$ 	& $.0038-.0094$	& $ .75-1.3$ \\
\hline
$B_{\chi}$ 		& $1.5$		& $3.4$		& $9.5$		& $101$		& $1.9$		& $1.5$	\\
$\tilde{m}_{\bar{\chi}}^2 $ & $3.9$	& $3.5$		& $.24$		& $18$		& $4.7$		& $2.5$	\\
$h_{\lambda} \left\langle \chi \right\rangle   / \lambda $
 			& $2.3$		& $1.0$		& $11$		& $7.9$		& $4.6$		& $1.8$	\\
 $\frac{h_{\gamma}\left\langle \chi \right\rangle \gamma}{16\pi^2}$
 			& $.033$	& $.054$	& $.16$		& $1.9$		& $.036$	& $.032$\\
 \hline
 $\left\langle F_{\chi} \right\rangle$ estimate
 				& $10$		& $10$		& $25$		& $130$		& $10$		& $10$	\\
 $\frac{\gamma^{ia} |F_{\chi_a}|}{\tilde{m}_{sq}^2}$  estimate
			& $10^{-7}$	& $10^{-7}$	& $10^{-7}$	& $10^{-5}$	& $10^{-7}$ 	& $10^{-7}$ \\
 $\longrightarrow \; \; \bar{\theta}$
			& $10^{-11}$	& $10^{-11}$	& $10^{-11}$ 	& $10^{-8}$ 	& $10^{-11}$ 	& $10^{-11}$ \\
\hline
$\bar{\theta}$	(quark)
			& $10^{-14}$	& $10^{-13}$	& $10^{-11}$	& $10^{-9}$	& $10^{-13}$ 	& $10^{-13}$ 	\\
$\bar{\theta}$	(gluino)
			& $10^{-15}$	& $10^{-15}$	& $10^{-14}$	& $10^{-11}$	& $10^{-15}$ 	& $10^{-15}$ 	\\
\hline\hline
\end{tabular}


\clearpage


\begin{thebibliography}{99}

\bibitem{instanton}
A.A. Belavin, A.M. Polyakov, A.S. Schwartz, and Yu.S. Tyupkin,
Phys. Lett. {\bf 59B}, 85 (1975); V.N. Gribov (unpublished);
G. 't~Hooft, Phys. Rev. Lett. {\bf 37}, 8 (1976); Phys. Rev. {\bf D14},
3432 (1976);
R. Jackiw and C. Rebbi, Phys. Rev. Lett. {\bf 37}, 172 (1976);
C.G. Callan Jr., R.F. Dashen, and D.J. Gross, Phys. Lett. {\bf 63B},
334 (1976).
\bibitem{nedm}
V. Baluni, Phys. Rev. {\bf D19}, 2227 (1979);
R.J. Crewther, P. di Vecchia, G. Veneziano, and E. Witten,
Phys. Lett. {\bf 88B}, 123 (1979); {\it ibid.} {\bf 91B}, 487(E) (1980).
\bibitem{visax}
S. Weinberg, Phys. Rev. Lett. {\bf 40}, 223 (1978); F. Wilczek, {\it ibid.}
{\bf 40}, 279 (1978).
\bibitem{invisax}
J.E. Kim, Phys. Rev Lett. {\bf 43}, 103 (1979);
M. Shifman, A. Vainshtein, and V. Zakharov, Nucl. Phys. {\bf B166}, 493 (1980);
A. Zhitnitsky, Yad. Fiz. {\bf 31}, 497 (1980) [Sov. J. Nucl. Phys. {\bf 31},
260 (1980)];
M. Dine, W. Fischler, and M. Srednicki, Phys Lett. {\bf 104B}, 199 (1981);
M. Wise, H. Georgi, and S.L. Glashow, Phys. Rev. Lett. {\bf 47}, 402 (1981);
H.P. Nilles and S. Raby, Nucl. Phys. {\bf B198}, 102 (1982).
\bibitem{Weinberg}
S. Weinberg, Trans. N.Y. Acad. Sci. {\bf 38}, 185 (1977).
\bibitem{KapMan}
D.B. Kaplan and A.V. Manohar, Phys. Rev. Lett. {\bf 56}, 2004 (1986);
K. Choi, C. W. Kim, and W. K. Sze, {\it ibid.} 61, 794 (1988);
K. Choi, Nucl. Phys. {\bf B383}, 58 (1992),
\bibitem{PQ}
R.D. Peccei and H. Quinn, Phys. Rev. Lett. {\bf 38}, 1440 (1977);
Phys. Rev. {\bf D16}, 1791 (1977).
\bibitem{Raffelt}
G. Raffelt, Phys. Rep. {\bf 198}, 1 (1990), and references therein.
\bibitem{Turner}
M. Turner, Phys. Rev. Lett. {\bf 60}, 1797 (1988); Phys. Rep. {\bf 197},
67 (1990).
\bibitem{axcosmo}
J. Preskill, M. Wise, and F. Wilczek, Phys. Lett. {\bf 120B}, 127 (1983);
L.F. Abbott and P. Sikivie, {\it ibid.} {\bf 120B}, 133 (1983);
M. Dine and W. Fischler, {\it ibid.} {\bf 120B}, 137 (1983).
\bibitem{aPl}
R. Holman, S.D. Hsu, T.W. Kephart, E.W. Kolb, R. Watkins,
and L.M. Widrow, Phys. Lett. {\bf B282}, 132 (1992);
M. Kamionkowski and J. March-Russell, {\it ibid.}
 {\bf B282}, 137 (1992); S. Barr and D. Seckel, Phys. Rev. {\bf D46},
539 (1992); B.A. Dobrescu, {\it ibid.} {\bf D55}, 5826 (1997).
\bibitem{KLLS}
R. Kallosh, A. Linde, D. Linde, and L. Susskind, Phys. Rev. {\bf D52},
912 (1995).
\bibitem{axexpts}
For current experimental axion searches see {\it e.g.}:
K. van Bibber, in {\bf Trends in Astroparticle Physics}, eds. D. Cline and
R.D. Peccei (World Scientific, Singapore, 1992);
I. Ogawa, S. Matsuki and K. Yamamoto, Phys. Rev. {\bf D53}, R1740 (1996).
\bibitem{DLM}
M. Dine, R. Leigh and D. MacIntyre, Phys. Rev. Lett. {\bf 69}, 2030
(1992); K. Choi, D.B. Kaplan, and A.E. Nelson, Nucl. Phys. {\bf B391},
515 (1993).
\bibitem{NB}
A. Nelson, Phys. Lett. {\bf 136B}, 387 (1984); S. M. Barr,
Phys. Rev. Lett. {\bf 53}, 329 (1984); Phys. Rev. {\bf D30}, 1805
(1984); A. Nelson, Phys. Lett. {\bf 143B}, 165 (1984).
\bibitem{PW}
J. Polchinski and M. Wise, Phys. Lett. {\bf 125B}, 393 (1983);
see also, F. del Aguila, M.B. Gavela, J.A. Grifols, and A. Mendez,
{\it ibid.} {\bf 126B}, 71 (1983); W. Buchm\"{u}ller and D. Wyler,
{\it ibid.} {\bf 121B}, 321 (1982); E. Franco and M. Mangano,
{\it ibid.} {\bf 135B}, 445 (1984);
M. Dugan, B. Grinstein, and L. Hall, Nucl. Phys. {\bf B255}, 413 (1985).
\bibitem{BMS}
S.M. Barr and A. Masiero, Phys. Rev. {\bf D38}, 366 (1988);
S.M. Barr and G. Segr\`{e}, {\it ibid.} {\bf 48}, 302 (1993).
\bibitem{DKL}
M. Dine, R. Leigh, and A. Kagan, Phys. Rev. {\bf D48}, 2214 (1993).
\bibitem{GMSB}
M. Dine and A. Nelson, Phys. Rev. {\bf D48}, 1277 (1993); M. Dine,
A. Nelson, and Y. Shirman,  {\it ibid.} {\bf 51}, 1362 (1995);
M. Dine, A. Nelson, Y. Nir, and Y. Shirman,  {\it ibid.} {\bf 53}, 2658
(1996).
\bibitem{MMM}
K.S. Babu, C. Kolda, and F. Wilczek,
Phys. Rev. Lett. {\bf 77}, 3070 (1996);
S.~Dimopoulos, S.~Thomas, and J.D.~Wells,
Phys. Rev. {\bf D54}, 3283 (1996); Nucl. Phys. {\bf B488}, 39 (1997);
J.A.~Bagger, K.~Matchev, D.M.~Pierce, and R.~Zhang,  
Phys. Rev. {\bf D55}, 3188 (1997).
\bibitem{fx}
S. Dimopoulos, G.F. Giudice, and A. Pomarol,
 Phys. Lett. {\bf B389}, 37 (1996);
S. P. Martin, Phys. Rev. {\bf D55}, 3177 (1997).
\bibitem{Bor}
 R. Rattazzi and U. Sarid, Nucl. Phys. {\bf B501}, 297 (1997);
 % CERN-TH/96-349, hep-ph/9612464.
 F.M. Borzumati, Report No. WIS-96/50/Dec.-PH, hep-ph/9702307.
\bibitem{rev}
For  a recent review and more references on GMSB, see
G.F. Giudice and R. Rattazzi, hep-ph/9801271.
\bibitem{aspon}
P.H. Frampton and T.W. Kephart, Phys. Rev. Lett. {\bf 66}, 1666 (1991).
\bibitem{FN}
P.H. Frampton and D. Ng, Phys. Rev. {\bf D43}, 3034 (1991).
\bibitem{asponB}
A.W. Ackley, P.H. Frampton, B. Kayser, and C.N. Leung,
Phys. Rev. {\bf D50}, 3560 (1994).
P.H. Frampton and S.L. Glashow, {\it ibid.}
{\bf 55}, 1691 (1997).
\bibitem{FK}
P.H. Frampton and O.C.W. Kong, Phys. Lett. {\bf B402}, 297 (1997).
\bibitem{domwall}
P. Sikivie, Phys. Rev. Lett. {\bf 48}, 1156 (1982).
\bibitem{weakinfl}
L. Randall and S. Thomas, Nucl. Phys. {\bf B449}, 229 (1995).
\bibitem{singlet}
V. Barger, M.S. Berger, amd R.J.N. Phillips,
 Phys. Rev. {\bf D52}, 1663 (1995);
 Y. Takeda, I. Umemura, K. Yamamoto, and D. Yamazaki,
  Phys. Lett. {\bf B386}, 167 (1996).
\bibitem{epK}
S.A. Abel and J.-M. Fr\`ere, Phys. Rev. {\bf D55}, 1623 (1997).
See also
T. Goto, T. Nihei, and Y. Okada, {\it ibid.} {\bf 53}, 5233 (1996);
F. Gabbiani, E.  Gabrielli, A. Masiero, and L. Silvestrini,
 Nucl. Phys. {\bf B477}, 321 (1996).
\bibitem{mia}
See, for example, J.S. Hagelin, S. Kelley, and T. Tanaka,
Nucl. Phys. {\bf B415}, 293 (1994).
\bibitem{x2}
G.C. Branco and L. Lavoura,
Nucl. Phys. {\bf B278}, 738 (1986);
F. del Aguila, M.K. Chase, and J. Cortes,
{\it ibid.} {\bf B271}, 61 (1986);
F. del Aguila and J. Cortes, Phys. Lett. {\bf 156B}, 243 (1985).
\bibitem{4h}
M. Masip and A. Ra\v{s}in, Phys. Rev. {\bf D52}, 3768 (1995).
%\bibitem{ram}
%H. Arason, D.J. Casta\~no, B. Kesthelyi, S. Mikaelian, E.J. Piard,
%P. Ramond, and B.D. Wright, Phys. Rev. {\bf D46}, 3945 (1992);
%D.J. Casta\~no, E.J. Piard, and
%P. Ramond, {\it ibid.} {\bf D49}, 4882 (1994).
\bibitem{MV}
S. P. Martin and M.T. Vaughn, Phys. Rev. {\bf D50}, 2282 (1994).
\bibitem{RGE}
Y. Yamada, Phys. Rev.   {\bf D50}, 3537 (1994);
I. Jack and D.R.T. Jones, Phys. Lett. {\bf 333B}, 372 (1994).

\end{thebibliography}
\end{document}